%% file: main.tex
\documentclass[a4paper,fleqn]{cas-dc}

\usepackage[numbers]{natbib}
\usepackage{comment}
\usepackage{caption}  
\usepackage[vlined,linesnumbered,ruled,boxed]{algorithm2e}
\usepackage{tikz}
\usepackage{graphicx}
\usepackage{subfig} 
\usepackage{enumitem}
\usepackage[font=small,labelfont=bf]{caption}
\usepackage{float}
\usepackage{pdfpages}

\def\tsc#1{\csdef{#1}{\textsc{\lowercase{#1}}\xspace}}
\tsc{WGM}
\tsc{QE}
\tsc{EP}
\tsc{PMS}
\tsc{BEC}
\tsc{DE}

\newcommand{\eg}{{\it e.g., }}

\newcommand{\name}{Action Engine\xspace}

\begin{document}
\let\WriteBookmarks\relax
\def\floatpagepagefraction{1}
\def\textpagefraction{.001}

\shorttitle{Automatic Workflow Generation in FaaS Continuum}

\shortauthors{Akiharu Esashi.}

\title [mode = title]{Action Engine: Automatic Workflow Generation in FaaS}                      
\tnotemark[1,2]



%
\author[1]{Akiharu Esashi}
[type=editor,
                       orcid=0009-0005-6919-8435]


\ead{akiharuesashi@my.unt.edu}


\credit{Conceptualization, Data curation, Investigation, Methodology, Project administration, Resources, Software, Validation, Visualization,  Writing – original draft, Writing – review and editing}

\affiliation[1]{organization={University of North Texas (UNT)},
    city={Denton},
    postcode={76203}, 
    state={Texas},
    country={U.S.A.}}

\author[2]{Pawissanutt Lertpongrujikorn}
[orcid=0009-0003-4106-2347]
\ead{pawissanuttlertpongrujikorn@my.unt.edu}
\credit{Conceptualization, Writing – original draft, Writing – review and editing, }
\affiliation[2]{organization={University of North Texas (UNT)},
    city={Denton},
    postcode={76203}, 
    state={Texas},
    country={U.S.A.}}

\author[3]{Shinji Kato}

\ead{shinjikato@my.unt.edu}

\credit{Software}
\author[4]{Mohsen Amini Salehi}
[orcid=0000-0002-7020-3810]

\ead{mohsen.aminisalehi@unt.edu}
\ead[URL]{https://hpcclab.org/}

\credit{Funding acquisition, Supervision}

\affiliation[3]{organization={University of North Texas (UNT)},
    city={Denton},
    postcode={76203}, 
    state={Texas},
    country={U.S.A.}}

\affiliation[4]{organization={University of North Texas (UNT)},
    city={Denton},
    postcode={76203}, 
    state={Texas},
    country={U.S.A.}}



\begin{abstract}
Function as a Service (FaaS) is poised to become the foundation of the next generation of cloud systems due to its inherent advantages in scalability, cost-efficiency, and ease of use. However, challenges such as the need for specialized knowledge, platform dependence, and difficulty in scalability in building functional workflows persist for cloud-native application developers. To overcome these challenges and mitigate the burden of developing FaaS-based applications, in this paper, we propose a mechanism called \name, that makes use of tool-augmented large language lodels (LLMs) at its kernel to interpret human language queries and automates FaaS workflow generation, thereby, reducing the need for specialized expertise and manual design. 
\name includes modules to identify relevant functions from the FaaS repository and seamlessly manage the data dependency between them, ensuring the developer's query is processed and resolved. Beyond that, \name can execute the generated workflow by injecting the user-provided arguments.
On another front, this work addresses a gap in tool-augmented LLM research via adopting an Automatic FaaS Workflow Generation perspective to systematically evaluate methodologies across four fundamental sub-processes. Through benchmarking various parameters, this research provides critical insights into streamlining workflow automation for real-world applications, specifically in the FaaS continuum. Our evaluations demonstrate that the \name achieves comparable performance to the few-shot learning approach while maintaining platform- and language-agnosticism, thereby, mitigating provider-specific dependencies in workflow generation. We notice that \name can unlock FaaS workflow generation for non-cloud-savvy developers and expedite the development cycles of cloud-native applications.
\end{abstract}


\input{highlights}

\maketitle

\input{intro}
\input{relatedwork}

\input{systemmodel}

\input{Evaluation}
\input{limitation}
\input{conclsn}

\section{Acknowledgment}

\noindent\textbf{Funding:}
This project is supported by National Science Foundation (NSF) through CNS Awards\# 2419588 and 2418188. We also used Chameleon Cloud to conduct evaluations of this research.

\noindent\textbf{Declaration of Generative AI and AI-Assisted Technologies in the Writing Process.}
During the preparation of this work, the authors used OpenAI’s ChatGPT to grammar-checking and rephrasing. After using this tool, the authors reviewed and edited the content as needed and take full responsibility for the content of the publication.

\printcredits

\balance
\bibliographystyle{plainnat}
\bibliography{references}

\onecolumn  

\appendix
\section{Additional Information of Action Engine}
\subsection{Task Planner Prompt Template}
\begin{figure*}[h]
    \centering
    \caption{The prompt used in Task Planner to divide a user-provided query into actionable sub-tasks.  }
    \vspace{-5pt}
    \includegraphics[width=1\textwidth]{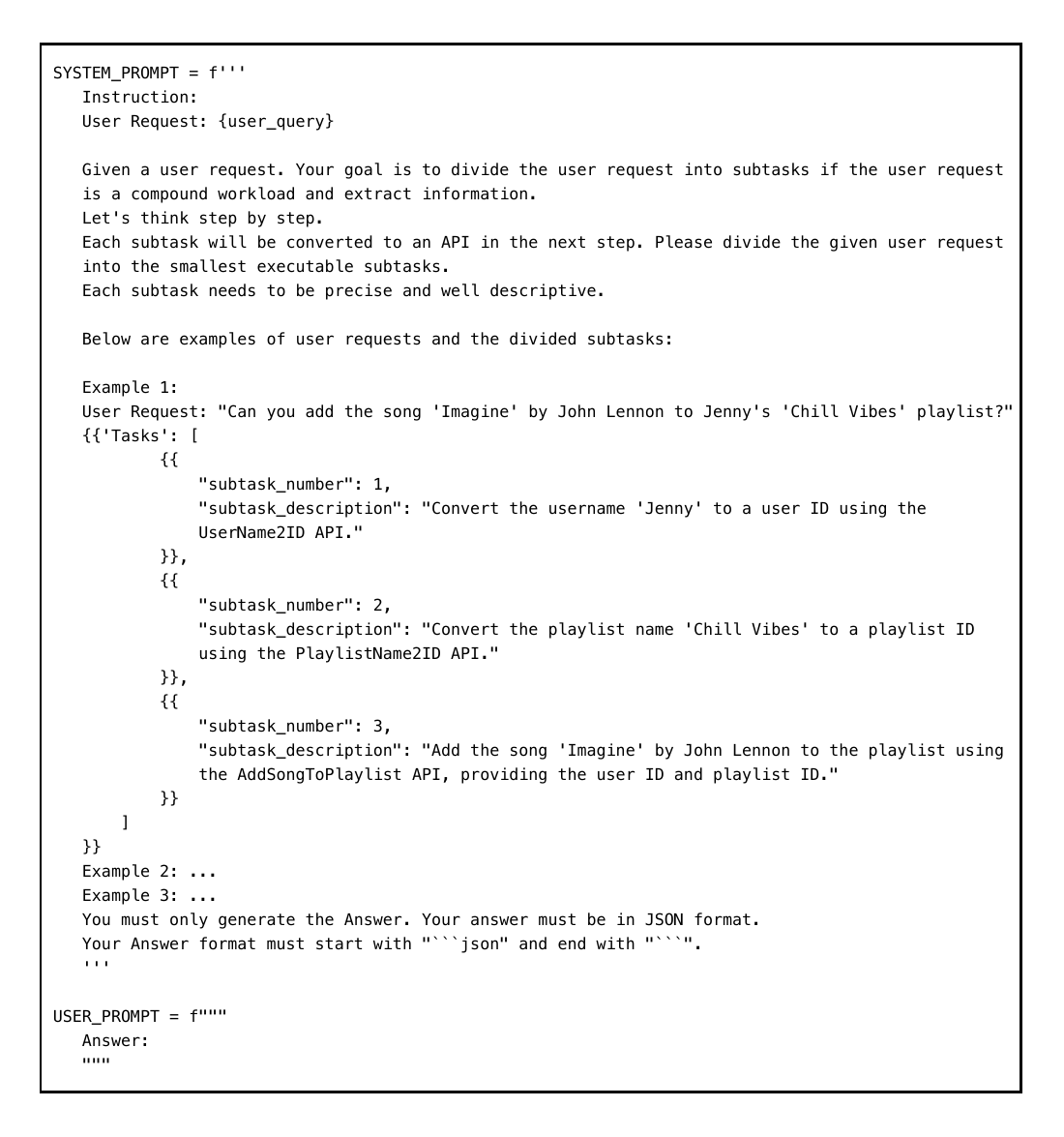}
    \label{AppendixA: Task Planner Prompt}

\end{figure*}

\newpage
\subsection{Function Selector Prompt Template}
\begin{figure*}[h]
    \centering
    \caption{The prompt used in Func Selector to select the most aligned function to a subtask description from candidate functions. }
    \vspace{-10pt}
    \includegraphics[width=1\textwidth]{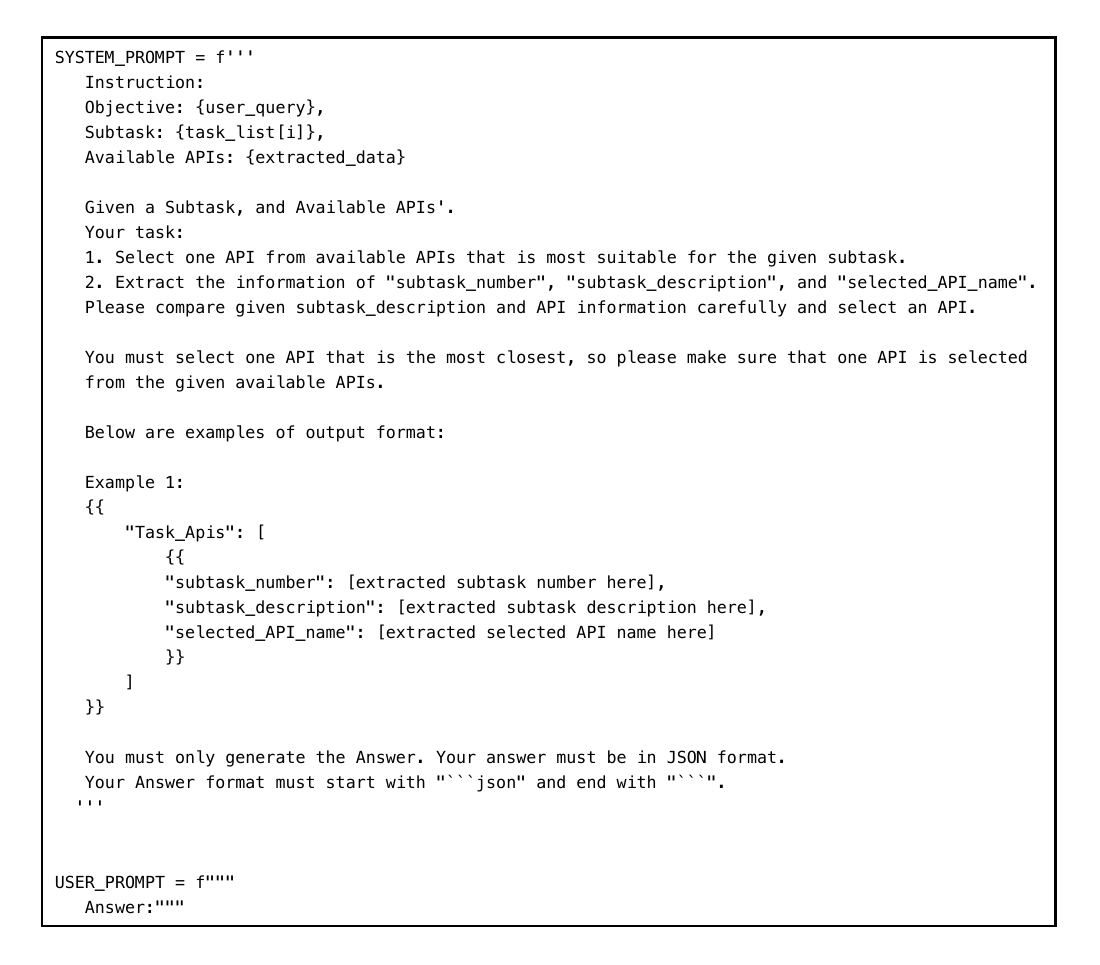}
    \label{AppendixA: FucnSelector Prompt}

\end{figure*}

\section{Additional Information for Experiments}
\label{AppendixB: fewshot}

\subsection{Zero-Shot and Zero-Shot CoT Prompt Templates}
\begin{figure*}[h]
    \centering
    \caption{The Zero-Shot (a) and Zero-Shot-CoT (b) prompt templates used in evaluations: In CoT prompt (b), the phrase "let's think step-by-step" is added.}
    \subfloat[Zero-Shot Prompt Template]{    \vspace{-5pt}\includegraphics[width=1\textwidth]{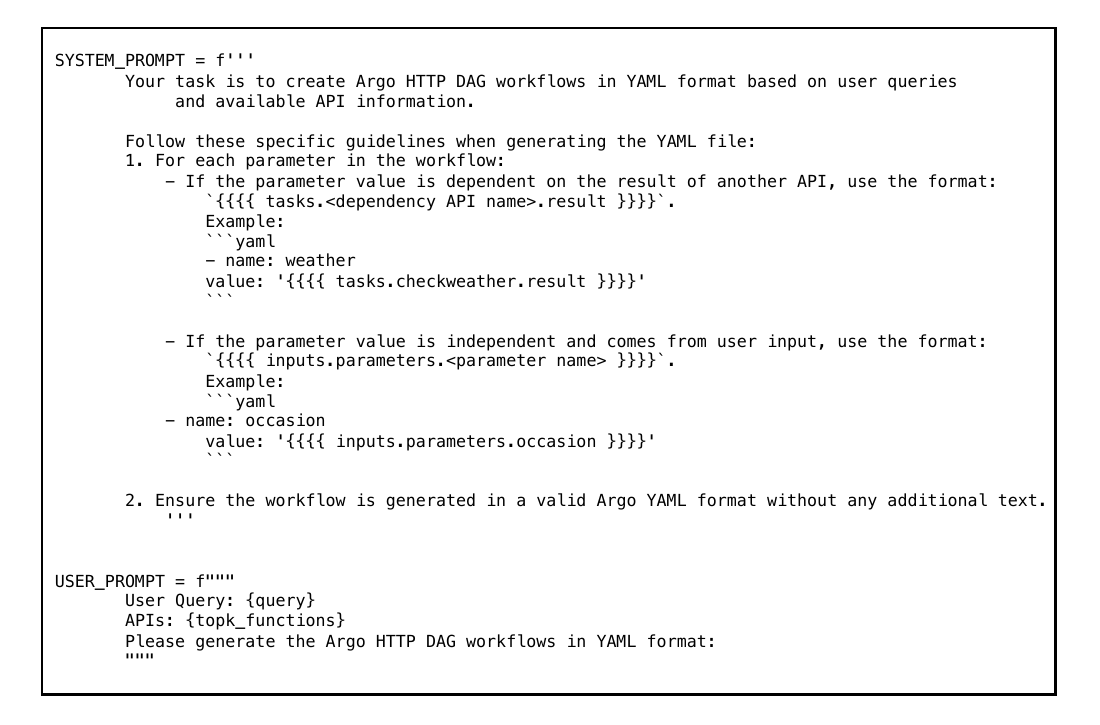}} 
    \vspace{-15pt}
    \subfloat[Zero-Shot-CoT Prompt Template]{    
    \includegraphics[width=1\textwidth]{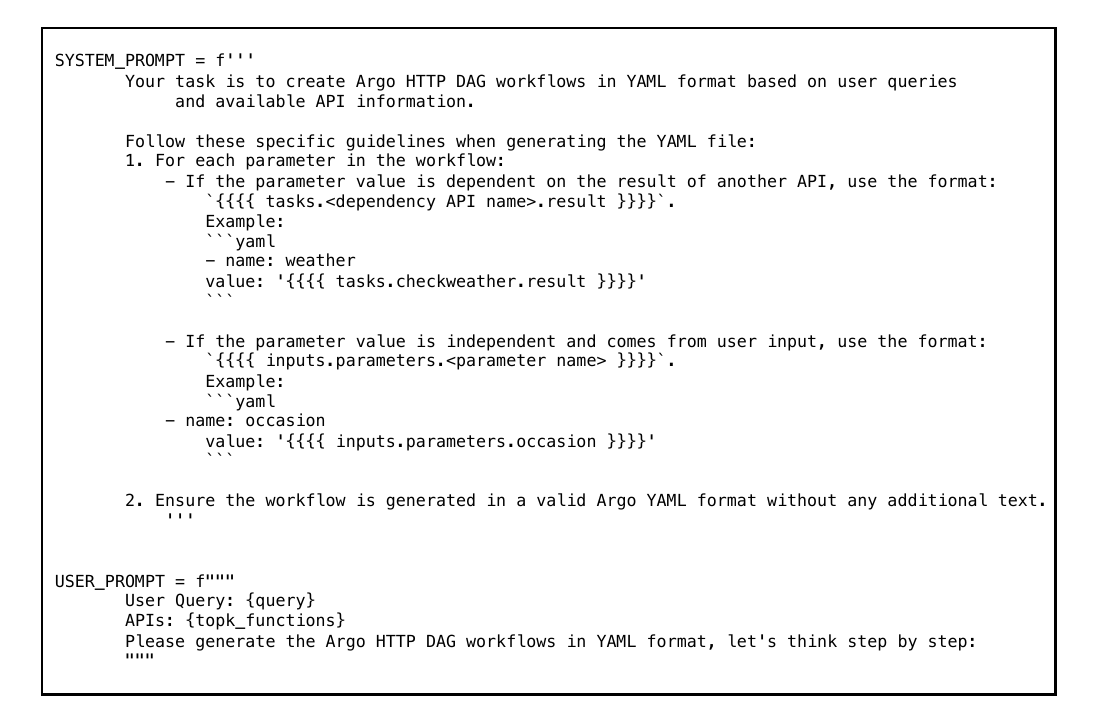}}
    
    \label{AppendixB: zeroshot and zeroshot-cot}
\end{figure*}\clearpage

\subsection{Few-Shot and Few-Shot CoT Prompt Templates}
\begin{figure*}[h]
    \centering
    \caption{
    Both Few-Shot and Few-Shot-CoT prompts templates used in evaluations follow the same structure with Zero-Shot and Zero-Shot-CoT format, respectively, with Few-Shot and Few-Shot-CoT incorporating an additional example of Argo Workflow generation demonstrations shown below. }
    \vspace{-10pt}
    \includegraphics[width=1\textwidth]{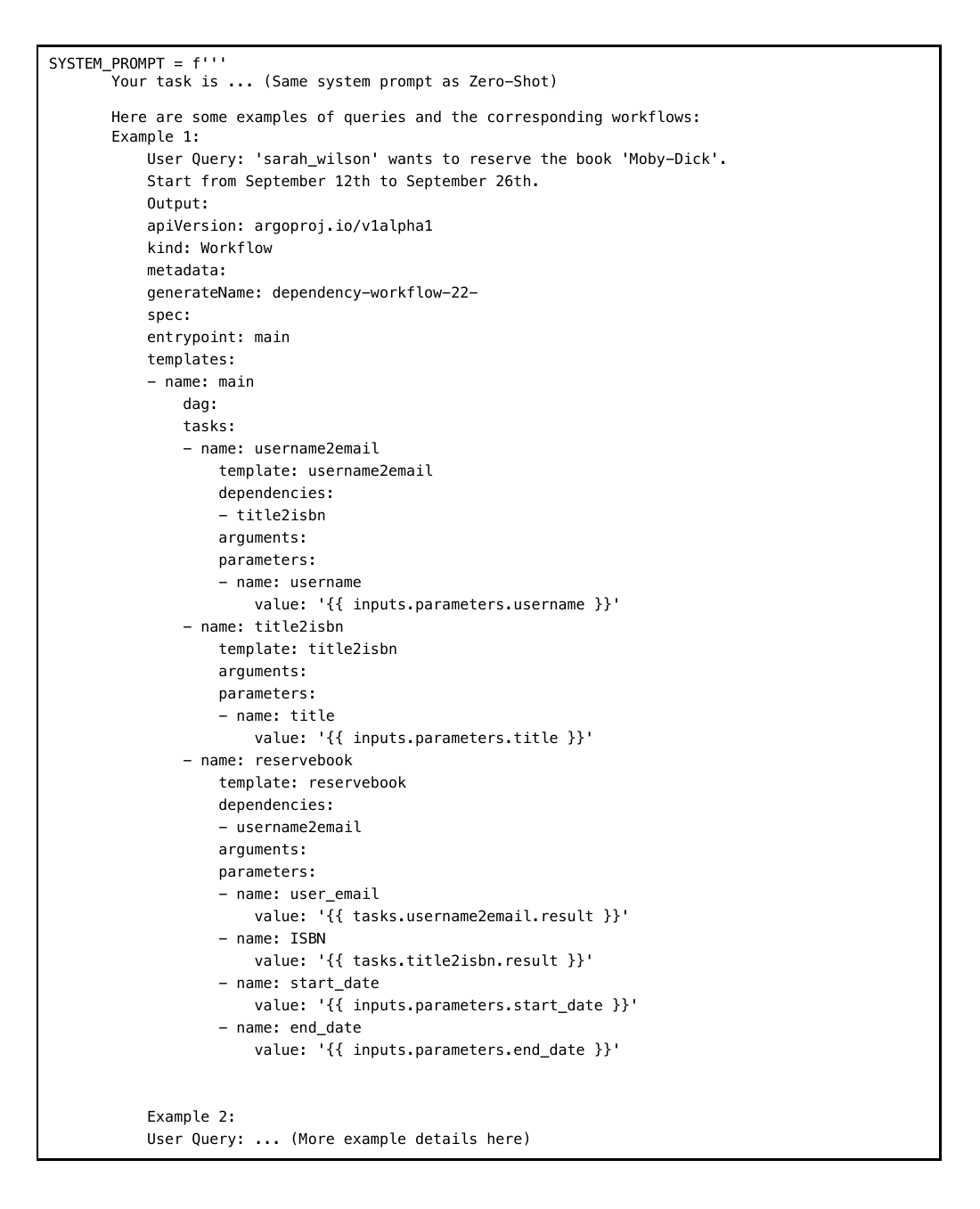}
    \vspace{-850pt}

\end{figure*}

\end{document}

%% file: highlights.tex
\begin{highlights}
\item Action Engine automates FaaS workflow generation using tool-augmented LLMs.
\item Enables platform-independent, language-agnostic FaaS workflows for cloud developers.
\item Reduces the need for manual coding and specialized expertise in serverless computing.
\item Ensures accurate function selection and data dependency management in workflows.

\end{highlights}

%% file: intro.tex
\section{Introduction}
\subsection{Function as a Service (FaaS)}
FaaS is a cloud computing paradigm that allows developers to execute code in response to events without managing the underlying infrastructure, providing scalability and cost efficiency \cite{chavitsurvey}. This model abstracts away the complexities of infrastructure management, thereby, enabling developers to concentrate on writing and deploying code triggered by specific events. As a result, FaaS enhances operational efficiency and accelerates cloud-native software development cycles. A key component of FaaS is its workflow orchestration, which enables the developer to orchestrate multiple functions to serve complex tasks by using high-level API. In cloud-native software development and Robotic Process Automation (RPA) \cite{hofmann2020robotic}, FaaS workflows are essential to automate tasks such as data processing pipelines.

FaaS is envisaged as the foundation of the next generation of cloud systems \cite{denninnart2021harnessing}, and major cloud providers have already adopted this paradigm, \eg AWS Lambda, Google Cloud Functions, and Azure Functions \cite{AWS-Lambda, Azure-functions, GoogleCloudFunctions}. These hyper-scalers also provide separate tools for workflow creation and orchestration, \eg AWS Step Functions, Google Cloud Composer, and Azure Durable Functions \cite{AWS-step-functions, Azure-Durable-Functions, GoogleCloudComposer}. 

\subsection{Challenges of FaaS Workflow Tools}

Despite the advancements in FaaS platforms, developers still face several challenges that make developing FaaS workflows difficult and time-consuming. In large software companies with multiple development teams, application workflows often require the integration of various functions from different teams. As a result, developers must understand and utilize cross-team functions, which can be both challenging and time-consuming. We summarize these challenges as follows:

\noindent\textbf{Specialized Knowledge Requirement:}
    Designing and implementing FaaS workflows is platform-specific and needs domain knowledge. For instance, AWS Step Functions demand familiarity with Amazon States Language (ASL) \cite{AmazonStatesLanguage}, and Google Cloud Composer necessitates knowledge of Python and Apache Airflow \cite{ApacheAirflow}. These steep learning curves are significant barriers for new developers in serverless cloud computing.

\noindent\textbf{Scalability Challenge:}
    Effective scaling of FaaS workflows demands a comprehensive understanding of existing functions in the repository to avoid redundancy and system flaws. As the function repository of an organization grows, it becomes more challenging for the developers to learn and appropriately utilize relevant functions to effectively create new workflows. 

\noindent\textbf{Provider Dependence:}
    FaaS applications are tightly coupled with provider-specific APIs, tools, and orchestration frameworks, introducing a significant platform dependence. Although there have been some efforts~\cite{khochare2023xfaas,ristov2024code} to enable a single codebase to be deployed across multiple platforms, there is currently no widely accepted standard for workflow definitions on any platform.  Consequently, workflow orchestrators still require platform-specific formats to define workflows. This reliance on provider-specific ecosystems restricts portability, complicates multi-cloud adoption, and raises development overhead when migrating or managing workflows across different cloud providers.
    
\noindent\textbf{Increasing Development Time:}
    Current FaaS workflow development relies heavily on manual coding, which is inefficient for large FaaS repositories. This reliance significantly increases the development time, making it challenging to adapt and respond quickly to new or changed requirements.

To address these challenges, we propose to use an LLM-based technology to help generate FaaS workflows. This will accelerate the development cycle and reduce the burden on developers who need to learn numerous functions, ultimately making them more productive.  

\subsection{LLM-Based FaaS Workflow Generation}
\begin{figure} [t]
    \centering
    \captionsetup{font=small,labelfont=bf} 

    \includegraphics[width=0.8\linewidth]{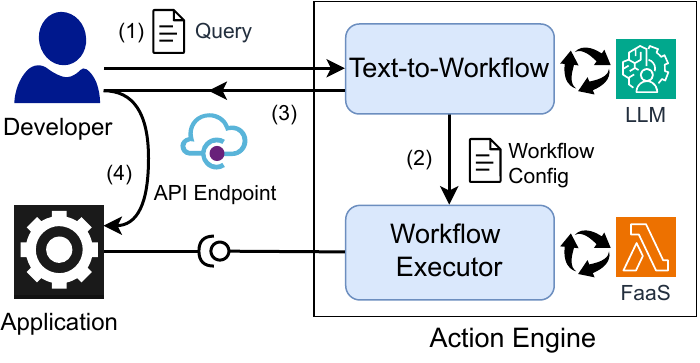}
    \caption{The high-level overview of \name. The developer submits the workflow description to \name to automatically generate the workflow and return the ready-to-use API endpoint backed by the FaaS platform and workflow orchestrator.}
    \label{fig:high-level}
\vspace{-10pt}
\end{figure}

The rapid advancements in Large Language Models (LLMs) have demonstrated remarkable capabilities in understanding and generating human-like text. More recently, LLMs have evolved to interact with external tools, called tool-augmented LLMs \cite{toollama_toolbench}, hence, are able to address complex tasks beyond text generation. 
Motivated by this emerging paradigm, we propose applying tool-augmented LLMs to address the unique challenges of FaaS workflow generation through our system, called \name and is shown in Figure~\ref{fig:high-level}. A solution to build FaaS workflows requires dynamic orchestration, scalability, and seamless integration across various functions. The conventional FaaS workflow generation approaches are limited by steep learning curves, platform dependencies, and the inefficiencies of manual workflow design. In contrast, \name leverages the advanced capabilities of tool-augmented LLMs to interpret natural language instructions, interact with FaaS functions, and automate the creation of workflows, thereby, effectively addressing these barriers.

Despite notable advancements in tool-augmented LLMs, the research remains nascent. Existing approaches often grapple with broad and inconsistent workflow definitions, platform, and language dependencies, and inadequate handling of data dependencies. These limitations make it particularly challenging to adapt tool-augmented LLMs to the FaaS workflow generation problem.
To overcome these challenges, in this paper, we design Action Engine, an end-to-end solution for automatic FaaS workflow generation. Action Engine builds on foundational concepts observed in current methods for tool-augmented LLMs, including tool planning, tool selection, tool invocation, and response generation. However, recognizing the unique demands of FaaS, we extend these concepts to encompass the platform-agnostic and dependency management features that are critical to FaaS workflows.

Upon receiving the user's (developer's) query, \name generates a directed acyclic graph (DAG), ensuring accurate dataflow across functions and in a FaaS platform-agnostic manner. By incorporating a data dependency module and an automated compilation process, Action Engine can transform the abstract (high-level) workflow descriptions into an executable FaaS workflow, while providing the user with an intuitive API endpoint to invoke the generated workflow. Specifically, the key contributions of this paper are as follows:


\noindent\textbf{(A) Architecture for Automatic FaaS Workflow Generation:} We designed and open-source\footnote{\href{https://github.com/hpcclab/action_engine}{https://github.com/hpcclab/action\_engine}} framework that leverages Tool-Augmented LLMs to generate FaaS workflows automatically. This framework reduces the need for specialized knowledge and manual intervention, thereby, mitigating the burden of workflow generation, management, and execution.

\noindent\textbf{(B) Platform-Independent Workflow Generation:} The proposed architecture can generate platform/cloud-independent workflows that can be later compiled to any underlying cloud platform or even potentially to functions across multiple clouds. 
    
\noindent\textbf{(C) Analysis of Tool-Augmented LLMs' Shortcomings:} We evaluate and analyze the limitations of tool-augmented LLMs for automatic FaaS workflow generation and identifying challenges such as inconsistent workflows, platform dependencies, and insufficient evaluation baselines that hinder adaptability to the FaaS environments.

These contributions collectively mitigate the burden and accelerate the development of FaaS workflows that can ultimately pave the way for democratizing cloud-native application development. The rest of the paper is organized as follows: 
Section~\ref{sec:background} presents an overview of existing approaches to workflow generation with LLMs and highlights their limitations. Section~\ref{sec:design} introduces the design of Action Engine. Next, our solutions to the challenges of \name are elaborated in Sections~\ref{sec:fn-id} and~\ref{sec:wf-gen}. Section~\ref{sec:evaluation} evaluates the performance of Action Engine from various dimensions. Section~\ref{sec:limitation_futurework} discusses the limitations and future works. Finally, Section~\ref{sec:conclusion} concludes the paper with a summary of key findings.

%% file: relatedwork.tex
\section{Background and Related Works}
\label{sec:background}

\subsection{Tool-Augmented LLM}

Tool-augmented LLMs are an emerging area of interest aimed at enhancing the ability of LLMs to interact effectively with various tools, such as APIs, thereby, demonstrating their impressive capability to solve real-world problems. Formally speaking, Tool-Augmented LLMs refer to the process that ``aims to unleash the power of LLMs to effectively interact with various tools to accomplish complex `tasks'`` \cite{toolnet,  toollama_toolbench}. Such LLMs harness the power of comprehending human language and selecting tools, all within a defined reasoning process, without the need for human intervention.

The definition of \textit{tools} often remains vague and inconsistent across different studies. Many notable existing works in this field employ the tool as a set of external functionality accessible by APIs \cite{apibank,  liu2023controlllm, gorillaLLM,toollama_toolbench, toolformer_meta, hugginggpt, restgpt, craft_tool_retrieval}. We adopt this definition, treating tools as APIs. In the context of FaaS, we utilize FaaS functions, which are accessed through APIs. Consequently, throughout this paper, the terms ''function'' and ''API'' will be used interchangeably, both referring to tools within the tool-augmented LLMs paradigm.


The process of tool learning typically consists of a four-stage process: \textit{task planning}, \textit{tool selection}, \textit{tool invocation}, and \textit{response generation}  \cite{toollearningsurvey}. LLMs begin the \textit{task planning} phase by analyzing user queries and decomposing them into actionable sub-tasks. The \textit{tool selection} phase then identifies the most appropriate tools to address these sub-tasks. Next, the \textit{tool invocation} phase retrieves the necessary parameters from the inputs and executes the corresponding functions. Finally, in the \textit{response generation} phase, the LLM synthesizes the outputs from these external tools to generate a coherent, context-aware response for the user. This structured process has been widely adopted in numerous studies on tool-augmented LLMs \cite{tptu2023, hugginggpt, taskbench, restgpt}, proving the significant potential of LLMs when combined with external tools.

\subsection{Automatic Workflow Generation with LLMs}

Existing research has explored various methods for integrating LLMs into automated workflow generation, demonstrating their potential to streamline complex tasks.
Notable examples include FlowMind \cite{flowmind} that utilizes LLMs to generate Python-based workflows via mapping user requests to financial APIs, and dynamically modifying workflows based on the users' real-time feedback. Similarly, AutoFlow \cite{autoflow} employs in-context learning and fine-tuning augmented with reinforcement learning to generate conceptual workflows.
While these approaches highlight the promise of LLMs for workflow generation, they primarily focus on conceptual designs and single-instance workflows. This limited scope leaves key challenges, such as reproducibility and language independence—largely unaddressed. Our paper aims to bridge these gaps.

\subsection{Limitations of LLM-based Workflow Generation}
\label{subsec:limitation_of_awg}
While tool-augmented LLMs demonstrate a significant potential for automating workflow generation, several key limitations hinder their application in FaaS workflow generation. These challenges can be categorized into the following three areas:

\noindent\textbf{(a) Broad and Inconsistent Workflow Definitions:}  
There is a lack of a consistent definition for workflows. Some studies interpret workflows as sequences of Python functions or scripts \cite{flowmind}, whereas others treat them as conceptual representations in natural language \cite{autoflow}. This inconsistency complicates efforts to standardize or extend existing methods, particularly for FaaS workflows.

\noindent\textbf{(b) Language Dependence:}  
Existing approaches predominantly rely on specific programming languages or utilize platform-specific APIs. Such dependency introduces challenges, such as reduced portability, increased maintenance overhead, and potential compatibility issues once the APIs or the runtime environments evolve. Additionally, LLMs can potentially produce syntactically or semantically invalid outputs, particularly for less popular platforms or languages.

\noindent\textbf{(c) Inadequate Data Dependency Handling:}  
Current automatic workflow generation methods struggle with effectively managing data dependencies in FaaS environments. While many systems support single-instance workflows, real-world FaaS deployments require explicit and deterministic data flow to ensure reliable and reproducible execution. Existing approaches often rely on implicit reasoning to establish data connections between functions, resulting in tightly coupled workflows that are difficult to reuse. For example, a workflow hard-coded with a specific username  fails in processing data for a different user---limiting both flexibility and generalizability of the solution.

To address this, the FaaS workflow generation solution must distinguish between \textbf{Data Dependencies} and \textbf{User-Defined Inputs}. The former refers to the explicitly defined connections between function outputs and inputs to ensure seamless data flow between workflow steps, whereas the latter is about the parameters dynamically provided at runtime---enabling workflows to adapt to varying contexts without requiring any structural modifications.

By embracing these principles and eliminating implicit reasoning, LLMs can generate reusable, adaptable, and reproducible FaaS workflows, hence, establishing robust automation solutions.

%% file: systemmodel.tex
\section{\name}
\label{sec:design}



The limitations of existing solutions, discussed in Section~\ref{subsec:limitation_of_awg}, 
highlights the need for a tailored approach to address the specific requirements of FaaS workflows. As such, we propose \name, an end-to-end solution designed specifically for FaaS workflow automation. Action Engine builds upon the foundational practices of tool-augmented LLMs while adapting them to the unique needs of FaaS workflows, namely platform independence, structured data flow, and reproducibility. Recall that tool-augmented LLM workflows follow a four-stage process: \textit{task planning}, \textit{tool selection}, \textit{tool invocation}, and \textit{response generation}. While effective for general-purpose applications, these stages are not held for the complexities of FaaS workflows that require precise data dependency management and language-agnostic representations.

\begin{figure} [t]
    \centering
    \includegraphics[width=\linewidth]{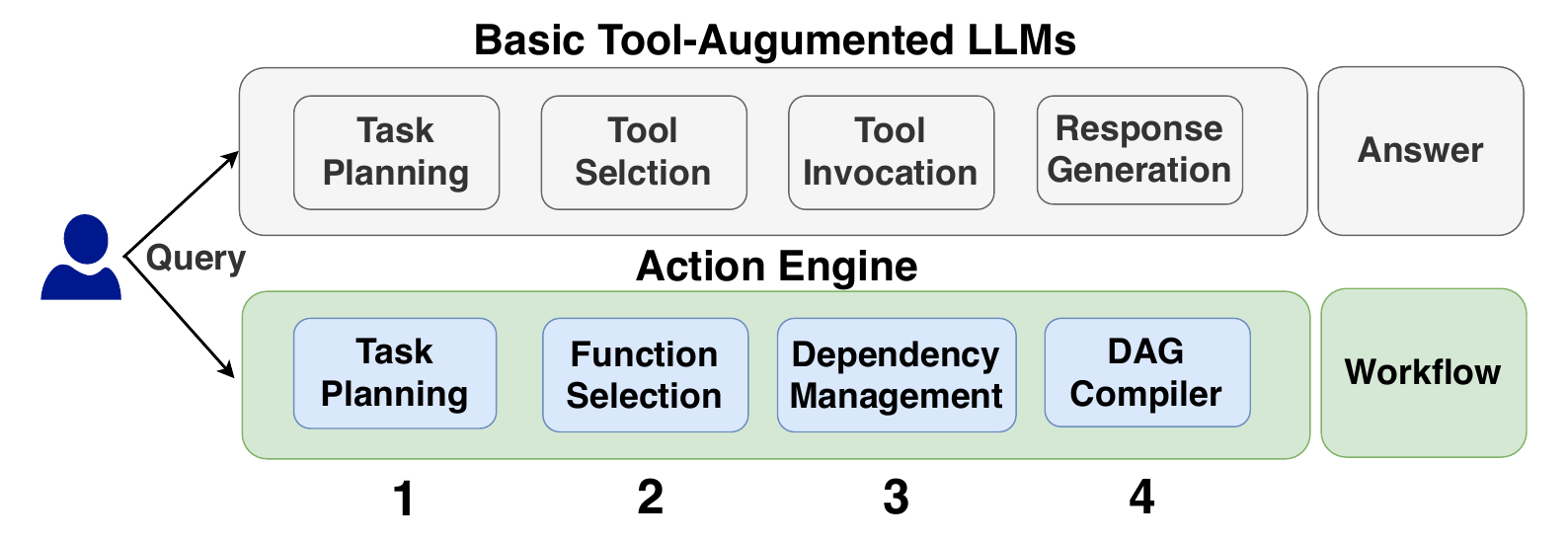}
    \caption{Comparison between the four standard steps of tool-augmented LLMs vs the proposed  processes for the FaaS workflow automation.}
    \label{fig:4process}
    \vspace{-10pt} 
\end{figure}

\name redefines these four foundational stages to create a scalable solution tailored to FaaS workflows, as shown in Figure~\ref{fig:4process}:

\noindent\textbf{Abstracting Workflows as DAG.} Workflows are represented as DAGs instead of being tied to a particular programming language. This abstraction ensures language neutrality and compatibility across diverse FaaS platforms (provided by various cloud providers) via decoupling workflow representation from specific programming environments.

\noindent\textbf{Task Planning and Tool Selection.} The \textit{task planning} process remains unchanged, dividing user queries into actionable sub-tasks. However, tools are mapped to FaaS functions, ensuring scalability and adaptability across multiple cloud platforms.

\noindent\textbf{Data Dependency Management.}
\name replaces conventional \textit{tool invocation} process with a data dependency management module to ensure explicit and deterministic data flow. By defining clear input-output relationships, it eliminates implicit reasoning and enables reusable workflows.     

\noindent\textbf{Platform-Specific Workflow Compilation.} Instead of \textit{response generation}, Action Engine introduces a dedicated compiler module to translate the platform-neutral DAG into formats compatible with specific FaaS platforms. This step ensures compatibility for seamless deployment and execution, addressing platform-dependency limitations.

Through these adaptations, Action Engine bridges the gap between existing tool-augmented LLM practices and the specific requirements of FaaS workflows. 
The following section elaborates the architectural design Action Engine, illustrating how these adaptations are realized in practice. This includes defining DAG representation, managing data dependencies, and compiling platform-specific configurations to create executable workflows.

\subsection{Action Engine Architecture}

\begin{figure*}[t]
    \centering
    \vspace{-10pt} 
    \includegraphics[scale=0.6]{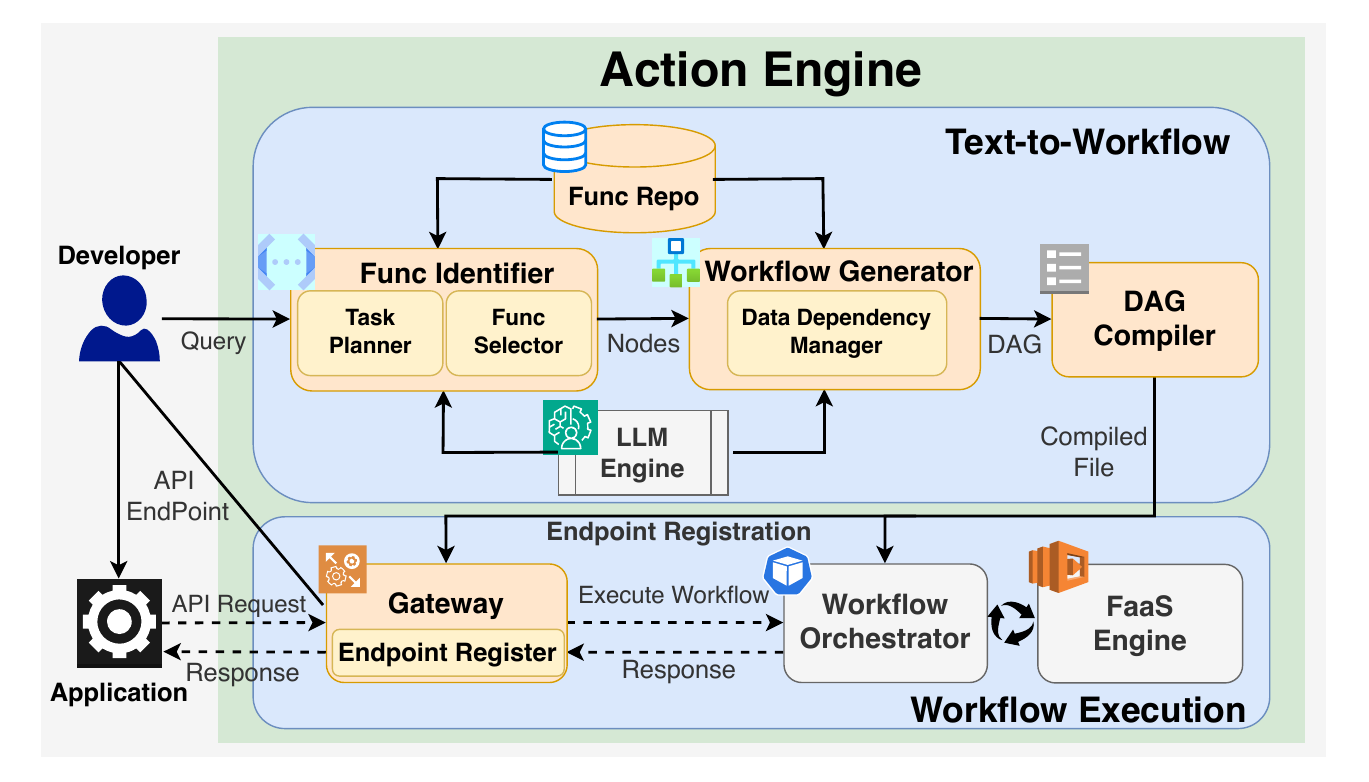}
    \vspace{-10pt} 
    \caption{Overview of the Action Engine architecture for FaaS workflow generation and execution. The system comprises two main modules: (1) Text-to-Workflow, which interprets user queries and constructs platform-agnostic DAGs through components such as LLM Engine, Function Identifier, Workflow Generator, and finally DAG Compiler turns it into the platform-specific workflow. (2) Workflow Execution, which deploys and executes the generated workflows using a Gateway, a Workflow Orchestrator, and a FaaS Engine. This layered design enables language-neutral, scalable FaaS orchestration with minimal developer effort.}
    \label{fig:architecture}
    \vspace{-15pt} 
\end{figure*}

Our approach to construct the FaaS workflow is to first define the DAG nodes, where each node consists of a sub-task derived from the user query and its corresponding identified function. The edges between nodes are identified through a ``data dependency'' approach that ensures a correct data flow between functions. Finally, we use a platform-specific configuration compiler to adapt the generated DAG into an executable workflow.

The overall architecture of \name is shown in Figure~\ref{fig:architecture}. As we can see in this figure, Action Engine consists of the following two modules:
\begin{enumerate} [leftmargin=*, noitemsep, topsep=0pt]
    \item \textbf{Text-to-Workflow}: Transforming the user's query into an executable workflow configuration. The generated workflow is assigned an API endpoint for the developer to call.
    \item \textbf{Workflow Execution}: Executing the workflow generated by the text-to-workflow module using its API endpoint.  
\end{enumerate}

\subsection{Text-to-Workflow Module} 
\label{text-to-worflow-each-components}
We design this module with the following five components:

\textbf{LLM Engine} processes internal prompts generated by the Func Selector and Workflow Generator components. The architecture is designed to support both open-source LLMs available on Hugging Face and closed-source models such as OpenAI's GPT series. We employ in-context learning \cite{in-context-learning} over fine-tuning due to the dynamic nature of FaaS, where the function repository can keep expanding with more new functions. Fine-tuning models typically require retraining on a fixed dataset, which can be inefficient and inflexible for the continuously changing environment of FaaS.

\textbf{Func Repo}\label{FuncRepo}  (Function Repository) is the centralized information storage for all available functions used within the system. Functions' information, including input and output parameters and functional descriptions, are stored as vector embeddings to facilitate efficient retrieval and selection during workflow generation. By embedding functions in a vector database, the system enables rapid similarity searches, significantly improving the efficiency of the function identification process (see more in Section~\ref{sec:fn-select}).
    
\textbf{Func Identifier} (Function Identifier) is to identify the appropriate function for each sub-task (node) of the DAG. This module operates in two key phases: Task Planning (Section~\ref{sec:task-plan}) and Function Selector (Section~\ref{sec:fn-select}).
    
\textbf{Workflow Generator} is to establish the data dependencies of each sub-task in the workflow, which are then used to construct the edges of the DAG. This module operates through three key processes: Topological Ordering (Section~\ref{sec:wf-gen}), Parameter Classification (Section~\ref{sec:wf-param-class}), and Data Dependency Construction (Section~\ref{sec:wf-dd}).
    
\textbf{Platform-Specific DAG Compiler} receives the generated DAG from the Workflow Generator and translates it into a specific workflow definition for the target workflow orchestrator. This component enables the objectives of language and platform independence from the workflow generation. The DAG is stored in a language-neutral format, allowing it to be compiled into definitions compatible with different workflow orchestrators by different DAG compilers. This component ensures that the generated workflows are not tied to a single platform and can be executed across various cloud environments without requiring manual reconfiguration.  Moreover, with the DAG structure, the compiler can easily identify independent sections of the workflow and optimize execution accordingly. Even when the target workflow orchestrator does not support DAG-based workflow definitions, such as AWS Step Functions~\cite{AWS-step-functions}, the compiler can strategically divide the workflow into multiple parts, facilitating parallel processing to improve performance. In contrast, some workflow orchestrators like Argo Workflow~\cite{argo-workflow} support DAG-based definitions and do not require this type of optimization.

\subsection{Workflow Execution}
To execute the generated workflow, we design the workflow execution module with the following three components:

\textbf{Gateway} is designed to provide the API endpoint for executing the generated workflow. Once the DAG compiler generates the workflow definition, the compiler module registers the workflow to the workflow orchestrator and the gateway to prepare the necessary data and resources. Next, the gateway creates an API endpoint and returns it back to the developer for integration with their application.
    
\textbf{Workflow Orchestrator} is a generic orchestrator for orchestrating multiple functions (task nodes in DAG) to be executed along with the FaaS engine. To execute the workflow, the application calls the API endpoint with the request data (user input). Upon receiving the request, the gateway forwards it to the workflow executor, which orchestrates the execution. The workflow is then decomposed into multiple sub-tasks, each executed as a separate function within the FaaS engine. After completing the workflow, whether successful or not, the orchestrator returns the result to the Gateway, which in turn forwards it back to the API caller.

\textbf{FaaS Engine} is a generic serverless platform that allows developers to run a function code as a scalable service effortlessly. \name uses the FaaS Engine in conjunction with the workflow orchestrator to execute the FaaS workflow.

\section{Func Identifier}
\label{sec:fn-id}
\subsection{Task Planner}
\label{sec:task-plan}

Task Planning involves breaking down user queries into actionable sub-tasks. This is achieved using in-context learning with carefully selected examples to ensure the appropriate granularity of task decomposition with respect to the available functions in the repository. For the user-provided query \( q \), the LLM function \( LLM\_plan \) submits \( q \), along with a set of example queries and their corresponding decompositions \( Q \), to the LLM in order to find the set of corresponding sub-tasks \( s_i \in S \).
\vspace{-3pt}
\begin{equation}
S = LLM\_plan(q, Q) = \{ s_1, s_2, \ldots, s_n \}
\label{eq:s}    
\end{equation}

As such, the LLM can apply the same decomposition principles, using the prompt template, detailed in Appendix A.1 (Figure \ref{AppendixA: Task Planner Prompt}), to convert any user-provided query into a set of manageable and executable actions, thereby facilitating the efficient execution of tasks.

\subsection{Function Selector}
\label{sec:fn-select}
Following Task Planning, the Function Selector is responsible for pairing each sub-task with the most relevant function from the function repository (Func Repo in Figure~\ref{fig:architecture}). To ensure that the selected functions are semantically aligned with subtask $s_i$, Function Selector uses cosine similarity between $s_i$ and the function embeddings to retrieve the top-k most relevant (i.e., highest cosine similarity scores) functions, denoted as $F^{s_i}_{1..k} = \{ f^{s_i}_1, f^{s_i}_2, \ldots, f^{s_i}_k \}$
, from the repository. 
At the \textit{function selection} stage, identifying incorrect selections is challenging before execution, except when function descriptions are clearly misaligned with the intended sub-task. To address this, based on our initial experiments, we apply a cosine similarity threshold of 0.7 to ensure that only function candidates with strong semantic alignment are considered. 

Subsequently, the LLM is utilized to select the most suitable function from the top-k retrieved functions (Appendix B, Figure \ref{AppendixA: FucnSelector Prompt}). To ensure the closest match, the LLM evaluates the semantic alignment and contextual appropriateness of each function relative to the sub-task. Specifically, the LLM chooses function \( f_i^* \) based on the parameters noted in Equation~\ref{eq:fi}.
\begin{equation}\label{eq:fi}
f_i^* = LLM\_select(q, s_i, F^{s_i}_{1..k})
\end{equation}

Let $N$ be the output of the Function Selector, a set of pairs in the form of \( (s_i, f_i^*) \) representing the DAG nodes where each $n_i\in N$ is defined as follows:
\begin{equation}
N = \{n_i \mid n_i = (s_i, f_i^*) \}
\label{eq:w}
\end{equation}

\section{Workflow Generator}
\label{sec:wf-gen}
The Workflow Generator module is designed to construct the data dependency (edges) between sub-tasks (nodes) of the workflow and form the DAG. Once the Function Selector creates the set of nodes $N$, the Workflow Generator performs topological ordering on the set of nodes $N$ to establish the correct execution sequence. This ordering is achieved by passing the list of sub-tasks to the LLM to semantically arrange the nodes. This process ensures that no node \( n_i \) is executed before its required preceding nodes are completed.

Next, for the function representing each node, the Workflow Generator classifies every input parameter to determine whether it depends on a direct user input or is an output generated by a previously executed node in the workflow. The classification result is used for constructing the data dependency that, in turn, forms the entire DAG.


\subsection{Parameter Classification} 
\label{sec:wf-param-class}
Let $\{p_{i1},p_{i2},\dots,p_{ik}\}$ be the set of parameters for the function of node $n_i$. Parameter Classification uses LLM to classify each parameter $p_{ij}$ as either a direct user input or an output from a previous sub-task. Let $\theta$ denote the set of semantic descriptions of output parameters from the previously executed nodes. The classification of parameter $p_{ij}$, denoted $t_{ij}$, is performed based on $\theta$, as shown in Equations~\ref{eq:class} and \ref{eq:class2}.
\begin{align}\label{eq:class}
t_{ij} &= LLM\_classify(p_{ij}, \theta) \\
&= \begin{cases}\label{eq:class2} 
Input & \text{if from direct user inputs} \\
Output(n_k) & \text{if from } n_k \text{ with } k < i
\end{cases}
\end{align}

\subsection{Data Dependency Construction}
\label{sec:wf-dd}
For each parameter \( p_{ij} \) classified as an output of a previous node, the Workflow Generator considers an edge $(n_{out},n_{in})$ as the data dependency in a DAG $E$. Let $n_s$ be a hypothetical node representing the starting node of the workflow holding the user inputs. For parameter $p_{ij}$ classified as an input, Workflow Generator
 creates an edge form  $n_s$ to $n_i$. Otherwise, for parameter $p_{ij}$ classified as an output of $n_k$, the system creates an edge from $n_k$ to $n_i$. We formally define $E$ as a set described in Equation~\ref{eq:edge}.
\begin{equation}\label{eq:edge}
\begin{split}
E = &\{(n_s,n_i)\mid t_{ij} = Input\} \\
& \cup\{(n_k,n_i)\mid t_{ij} = Output(n_k)\}
\end{split}
\end{equation}

The aforementioned procedure leads to the creation of a DAG defined as $DAG=<N,E>$.


%% file: evaluation.tex
\section{Performance Evaluation}
\label{sec:evaluation}
In this section, we assess the capabilities and performance of \name in automatic FaaS workflow generation by comparing them against existing methods in terms of accuracy and scalability.

\subsection{Experimental Setup}
\label{sec:evlt:setup}
\subsubsection{Dataset}
The evaluation employs an enhanced version of the Reverse Chain dataset \cite{reversechain}, specifically designed for compositional multi-tool tasks. This dataset includes 825 unique APIs spanning 20 diverse categories, resulting in a total of 1,550 labeled instances where each instance consists of a user query, a set of APIs with their descriptions and argument structures, and the corresponding ground truth workflow represented in a nested function-calling format.

The dataset is categorized into three levels of complexity based on the degree of API nesting required to resolve the queries. Level-1 involve two levels of API nesting and Level-2 extend to three levels of nesting. The most complex ones, Level-3, require more than four levels of nested API calls.

The ground truth for each instance specifies the intended API workflow in a structured, nested function calling format, similar to the example shown below:
\vspace{-2mm}
\begin{verbatim}
AddSongToPlaylist(
    user_ID=UserName2ID(user_name='Jenny'),
    playlist_ID=PlaylistName2ID(playlist_name='Chill Vibes'),
    song_name='Imagine'
)
\end{verbatim}
\vspace{-2mm}
This representation defines a sequence of dependent API calls, where arguments for one API are resolved by outputs from other APIs or user-provided information from the original query. Such structured formats effectively capture the relationships and dependencies between APIs.

The Reverse Chain dataset ensures explicit data dependency injection by structuring workflows where the outputs of one API directly serve as the inputs to the subsequent APIs. By providing ground truth workflows in a nested function-calling format and supporting different levels of complexity, this dataset aligns with the demands of evaluating FaaS workflow generation systems that require multi-step function orchestration.

During the pre-processing phase, we identified several inconsistencies in the dataset, and by addressing them, we affirmed its validity for the evaluations. Two major issues were observed: (a) invalid syntax and (b) illegal accessor usage. Illegal accessors occurred when the ground truth attempted to access attributes or fields of a function’s output that were not defined in the function’s behavior. For example, some workflows attempted to retrieve specific properties from a function when the output value was a string. Such mismatches cause logical errors in the ground truth data, which is possibly due to the use of LLMs by the Reverse Chain \cite{reversechain} to generate the dataset. To address these issues, we filtered out all the instances containing invalid accessors or syntax errors. After this cleanup, we conducted stratified random sampling to select a balanced subset of 300 examples, with 100 examples from each of the three complexity levels. This pre-processing ensured the dataset was both logically consistent and syntactically valid.

\subsubsection{Evaluation Metrics}

To measure the correctness of the FaaS workflow, we employ evaluation metrics that emphasize exactness in function selection, data dependencies, and topological correctness. Unlike general-purpose task evaluations that may tolerate semantic approximation, FaaS workflows require strict adherence to the prescribed logic and execution dependencies. Therefore, our evaluation framework prioritizes exact matches between generated workflows and the ground truth.

\textbf{Function Selection and Data Dependency.}
For function selection, correctness is determined via comparing the generated API names and their corresponding argument structures with the ground truth \cite{wu2024sealTool}. Any mismatches in function names or parameter values are treated as errors. For data dependencies, correctness is evaluated by verifying whether the generated workflows accurately reflect the input-output relationships specified in the ground truth. 

The metrics used to evaluate the results are \textit{precision, recall}, and the \textit{F1 score}, which are defined as follows:
\textbf{Precision:} The proportion of correctly predicted functions or data dependencies out of all predicted instances. Precision highlights the system's ability to avoid false positives in function or parameter generation.
\textbf{Recall:} The proportion of correctly predicted functions or data dependencies out of all ground truth instances. Recall measures the system's capability to capture all relevant elements.
\textbf{F1 Score:} The harmonic mean of precision and recall, providing a balanced evaluation metric. The F1 score ensures that both accuracy (precision) and completeness (recall) are equally considered in the assessment.

Unfortunately, the Reverse Chain dataset does not come with the function code or the online API that we can use to run the function. Hence, we are unable to evaluate the result of the generated workflow by verifying the execution results.

\textbf{Topological Order.}
To evaluate the correctness of the topological ordering of functions within a workflow, we utilize the \textbf{Longest Common Subsequence (LCS)} metric. LCS measures the similarity between the sequence of nodes in the generated workflow and the ground truth by identifying the longest subsequence that appears in both orderings. The topological order is considered correct if the generated sequence adheres to the logical dependencies required for workflow execution.
This metric is particularly suited for FaaS workflows, where the order of function execution directly impacts the correctness of the data flow and overall execution. A higher LCS score indicates better alignment between the generated and ground truth orderings.

\subsubsection{Baseline Methods}
To the best of our knowledge, this work presents the first system for automatic FaaS workflow generation. As there are no direct baselines for this specific task, we consider the broader area of the automatic code generation problem conditioned on a user request and candidate functions in YAML format. We evaluate our system against two state-of-the-art models. First, we use the \textbf{GPT-4o} model \cite{openai_gpt4o}, a state-of-the-art closed-source model from OpenAI. Second, we use \textbf{Qwen-Coder-32B-Instruct} \cite{qwen25_tech_report}, a fine-tuned model specifically designed for coding tasks, which has demonstrated coding capabilities comparable to GPT-4o \cite{bigcodebench}.

To comprehensively evaluate the performance of \name, we benchmark against the above models with the following four established in-context learning methods \cite{brown2020language}: 
\textbf{Zero-Shot} \cite{kojima2022large} refers to a scenario where the model performs a task without any prior examples, relying entirely on its pre-trained knowledge.
\textbf{Few-Shot} provides the model with a small number of task-specific examples to improve its ability to generate accurate responses.
\textbf{Zero-Shot Chain of Thought (CoT)} \cite{wei2022chain} extends zero-shot learning by prompting the model to break down its reasoning process step by step to enhance logical consistency.
\textbf{Few-Shot Chain of Thought (Few-Shot CoT)} \cite{wei2022chain} combines few-shot learning with the step-by-step reasoning approach of CoT to further improve the performance in complex decision-making tasks. The prompt template used during evaluation is provided in Appendix \ref{AppendixB: fewshot}.
    
In addition to these in-context learning baselines, we compare our solution against \textbf{Reverse Chain} \cite{reversechain}, an approach that constructs workflows in a top-down manner. Since the official implementation was not open-sourced, we adopted a third-party reproduction based on the original paper \cite{reversechain-github} and re-implemented our own version for fair comparison. Unlike our bottom-up approach, which begins by identifying dependencies and incrementally building workflows from input toward the goal, Reverse Chain starts with the final goal and works backward to determine the APIs and arguments needed to fulfill the required parameters. This top-down approach offers an alternative perspective on workflow generation, providing valuable insights into the strengths and limitations of different strategies.

We utilize GPT-4o as the underlying LLM model to generate Argo Workflow \cite{argo-workflow} for further evaluation of both \name and Reverse Chain. These methods are chosen for their relevance to tool planning and workflow construction and their demonstrated effectiveness in similar tasks. Argo Workflow was chosen for its native ability to define complex workflows as DAG and its open-source nature, which allows us to use it in our own cluster. This makes it a solid solution for workflow orchestration.


\begin{figure*}[h]
    \centering
    \vspace{-15pt}
    \subfloat[F1 Score - Function Selection]{\includegraphics[width=0.33\textwidth]{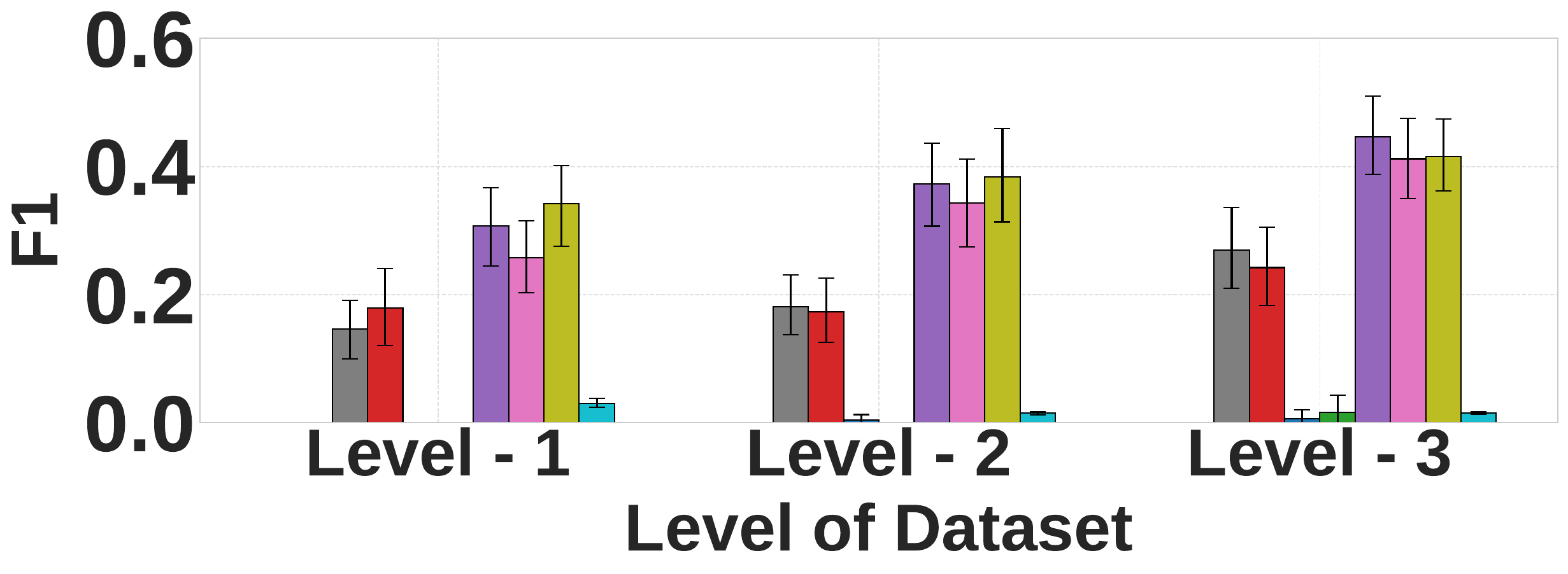}\label{fig:evlt:e2e:fs}}
    \subfloat[F1 Score - Data Dependency]{\includegraphics[width=0.33\textwidth]{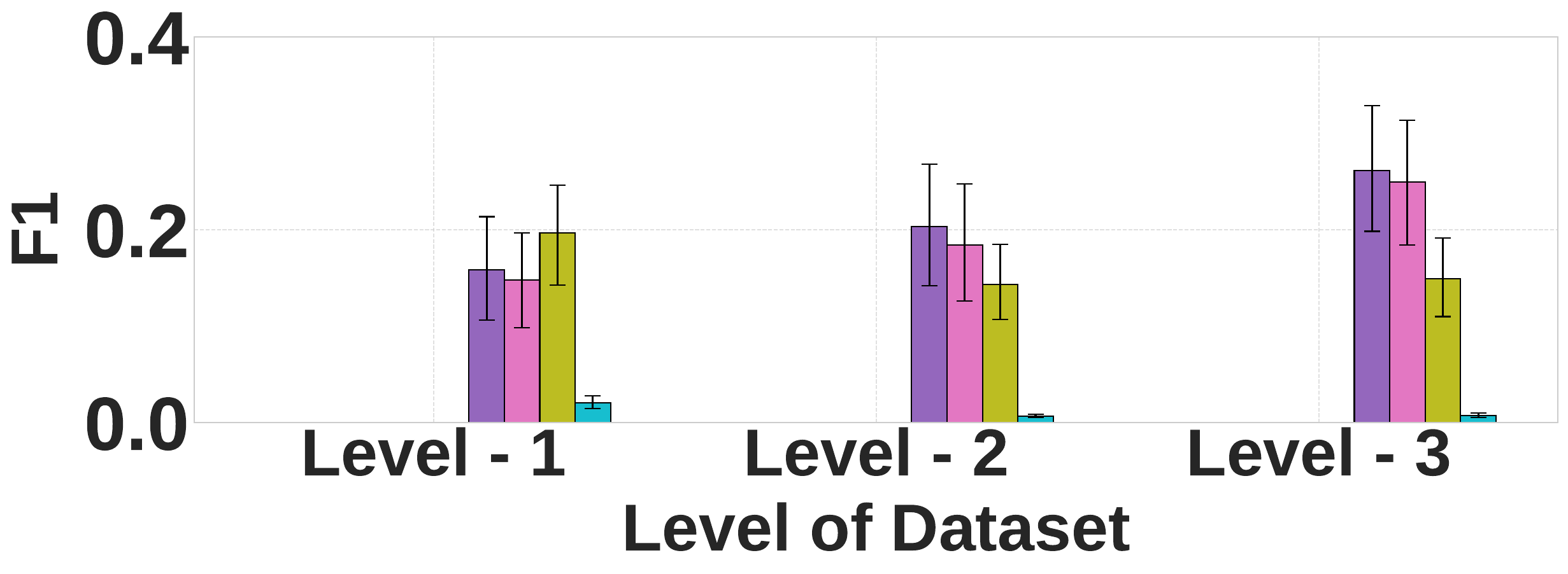}\label{fig:evlt:e2e:to}}
    \subfloat[LCS score - Topological Order]{\includegraphics[width=0.33\textwidth]{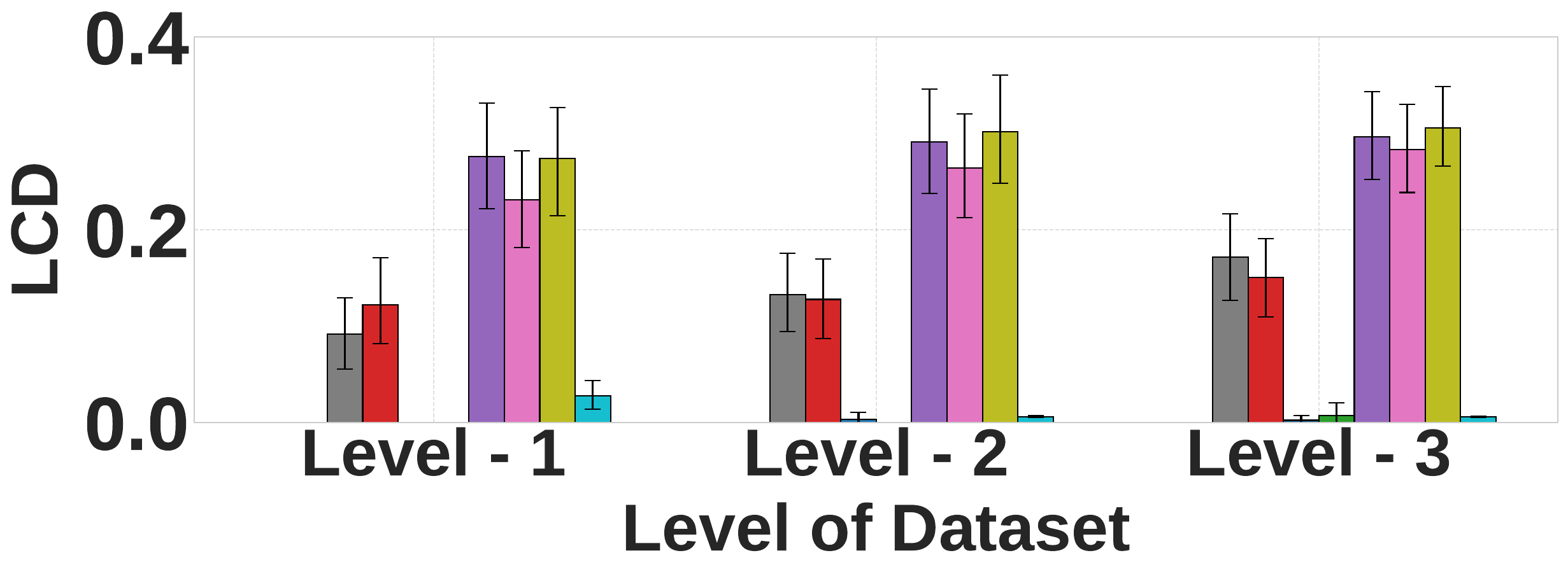}\label{fig:evlt:e2e:dd}}
    
    \vspace{-8pt}

    \subfloat{%
        \includegraphics[width=1\textwidth]{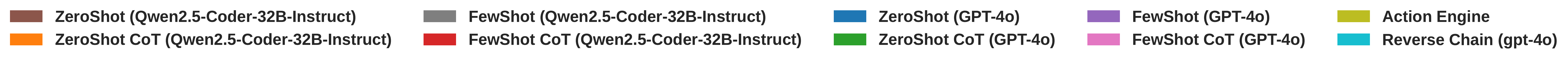}%
        \label{Figure4_legend}   }
\caption{Measuring workflow correctness of \name against the following baselines: Zero-Shot, Zero-Shot CoT, Few-Shot, Few-Shot CoT (each evaluated using both GPT-4o and Qwen-Coder-32B-Instruct), and Reverse Chain. These variations are evaluated against the same FaaS dataset and with queries categorized into three complexity levels.}
    \vspace{-15pt} 
    \label{fig:evlt:mainresults}
\end{figure*}

\subsection{Measuring Correctness of Generated Workflows}
\label{sec:evlt:mainresult}
As illustrated in all subfigures of \ref{fig:evlt:mainresults}, the workflow correctness improves as the workflow complexity increases, which is counterintuitive. This trend is evident across all evaluation metrics, with \name demonstrating a stable performance while other methods exhibit varying degrees of fluctuation. This phenomenon is further investigated in the ablation study in Section \ref{sec:eval:shortcomings-dataset}, where we analyze the impact of dataset ambiguities and the existence of multiple valid solutions in the dataset.

Despite its comparable coding ability to GPT-4o, Qwen-Coder-32B-Instruct provides less than half of the accuracy GPT-4o offers in performance across all metrics. This low performance suggests that strong code comprehension alone is insufficient for FaaS workflow generation that requires task decomposition and precise dependency understanding. These findings reinforce the importance of task-specific architectural choices in achieving a robust performance against query difficulty level.

Across various evaluation metrics, Action Engine demonstrates comparable accuracy to the GPT-4o-based FewShot and FewShot CoT baselines while offering the additional advantage of language- and platform-agnostic workflow generation.

For function selection (Figure~\ref{fig:evlt:mainresults}a), Action Engine achieves stable F1 scores of 34\%, 38\%, and 42\% across Levels 1, 2, and 3, indicating its robustness across different workflow complexities. In contrast, FewShot (GPT-4o) and FewShot CoT (GPT-4o) show gradual improvements, with F1 scores increasing from 30\% to 45\% and 26\% to 41\%, respectively. Similarly, in topological ordering, Action Engine consistently attains the highest scores, maintaining a stable LCS value of 30\% across all complexity levels.

As for the data dependency accuracy, although Action Engine achieves comparable results to FewShot (GPT-4o) and FewShot CoT (GPT-4o), it exhibits a slightly lower accuracy than them. This suggests that while the Action Engine is effective at function selection and workflow structuring, there is room for improvement in accurately capturing fine-grained data dependencies.

Across all metrics and complexity levels, ZeroShot, ZeroShot CoT, and the Reverse Chain exhibit severe limitations, achieving no more than 10\% in all three accuracy metrics. These results highlight fundamental deficiencies in these approaches even when utilizing state-of-the-art LLMs such as GPT-4o in handling tasks that require nuanced understanding, constructive planning, and precise dependency management. The Reverse Chain Method, in particular, fails to establish meaningful dependencies, further demonstrating its inadequacy for complex workflow generation. Given that the Reverse Chain results diverge from those reported in previous literature and theoretically should exhibit moderate accuracy, an ablation study (in Section \ref{ablation: reversechain}) further investigates the factors contributing to this discrepancy.

Among the five baseline methods, FewShot (GPT-4o), FewShot CoT (GPT-4o), and \name consistently achieve the highest accuracy across all evaluation metrics. Notably, \name surpasses other methods in both function selection and topological correctness while also achieving language-neutral workflow generation, making it the most versatile approach for automated workflow construction.

\begin{figure*}[h]    
    \centering
    \vspace{-10pt}
    \includegraphics[width=1\textwidth]{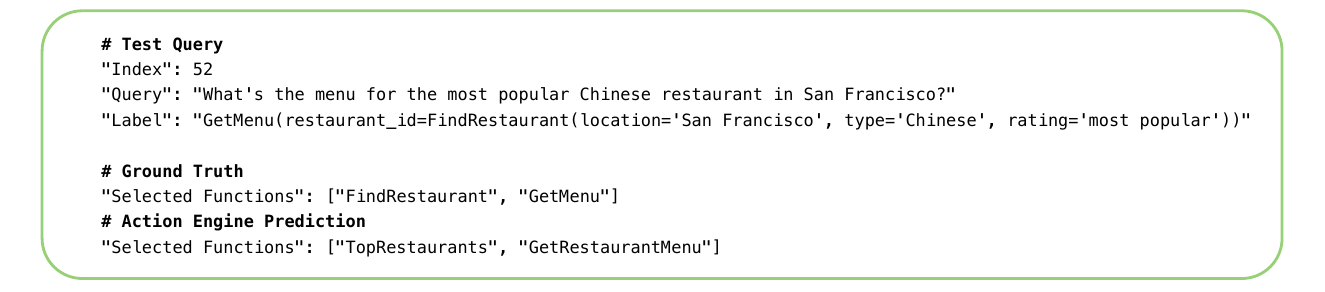}
    \vspace{-10pt}
    \caption{Example of multiple correct solutions in the Reverse Chain dataset. }
    \vspace{-10pt}
    \label{ablation:dataset} 
\end{figure*}

\subsection{Ablation Study}
\label{sec:evlt:ablation}
\subsubsection{Limitations of the Existing Evaluation Datasets}
\label{sec:eval:shortcomings-dataset}

As highlighted in the previous section, our results indicate a counterintuitive trend: accuracy appears to increase as the workflow complexity rises. Our analysis shows the reason for this behavior is that, as the workflow complexity increases, the number of decision points (nodes) also increases, which improves the overall chances of correctness. In simpler workflows with fewer nodes, errors have a greater impact because there are fewer opportunities for the workflow to align with the predefined ground truth. In contrast, more complex workflows contain a larger number of decision points, meaning that even if some selections are incorrect, the overall workflow is still more likely to be evaluated as correct. Increasing the number of nodes provides more opportunities for correct selections to contribute to the final accuracy, thereby, reducing the influence of isolated mistakes. However, traditional accuracy metrics often do not account for this probabilistic effect, leading to a misinterpretation of the relationship between complexity and accuracy.

While this may seem unexpected, further analysis reveals that this phenomenon stems from inherent ambiguities in LLM-generated datasets and the existence of multiple valid solutions. Two key factors that contribute to this evaluation challenge are as follows:
\begin{enumerate} [leftmargin=*, noitemsep, topsep=0pt]
    \item \textbf{Multiplicity of Valid Solutions:} Many workflows can achieve the same goal through different, yet equally valid, sequences of function selections. Traditional evaluation metrics, which rely on a fixed reference solution, often fail to account for this flexibility, leading to discrepancies in accuracy assessment.
    \item \textbf{Ambiguity in LLM-Generated Datasets:} The use of LLMs to generate datasets introduces inherent uncertainties. Simulated API tools frequently overlap in functionality, resulting in multiple plausible workflow configurations. As workflow complexity increases, defining a single ``correct'' reference solution becomes increasingly difficult, further complicating evaluation reliability.
\end{enumerate}
\vspace{2mm}

To further investigate this issue, we manually analyzed a subset of predictions from \name results and realized that many predictions, while differing in structure from the predefined ground truth, were functionally equivalent to the expected outcome (Figure~\ref{ablation:dataset}). This suggests that model-generated workflows are frequently penalized as incorrect despite successfully solving the task.
Upon closer examination of the selected functions in both the predictions and the predefined ground truth, we noticed that many functions had highly similar descriptions. This highlights an important challenge: within a large dataset of available functions, multiple functions often exhibit near-identical functionalities, creating multiple valid solutions. Consequently, even when ground truth data is manually curated, ensuring a singular, absolute reference path remains an unrealistic expectation.

This inherent ambiguity in evaluation necessitates a paradigm shift in assessment methodologies. Rather than relying solely on topological alignment, evaluation frameworks should incorporate a more functionally grounded approach. By shifting the focus from rigid workflow topology to correctness based on functional output, we can better capture the true effectiveness of the workflow generation models.

By adopting a more flexible and context-aware evaluation framework, we can improve the reliability of accuracy assessments, ultimately ensuring a fairer and more robust evaluation of workflow generation models. This shift is particularly crucial for real-world applications, where multiple valid workflows often exist, and correctness should be judged by workflow results rather than adherence to a single predefined path. Importantly, the presence of redundant or functionally similar APIs is not merely an artifact of the dataset but a realistic feature of practical FaaS environments. In deployed systems, overlapping functions are intentionally maintained to provide resilience, support interoperability, and enable fallback options. As such, evaluating workflow correctness purely based on topological alignment can obscure practical performance issues or overstate accuracy. These observations highlight the need for functionally grounded evaluation metrics that better reflect real-world utility.


\subsubsection{Analyzing the Impact of Top-Down Approach for Tool-Augmented LLMs}
\label{ablation: reversechain}

\begin{figure*}[h]
    \centering
    \vspace{-15pt}
    \subfloat[Precision - Function Selection]
    {\includegraphics[width=0.25\textwidth]{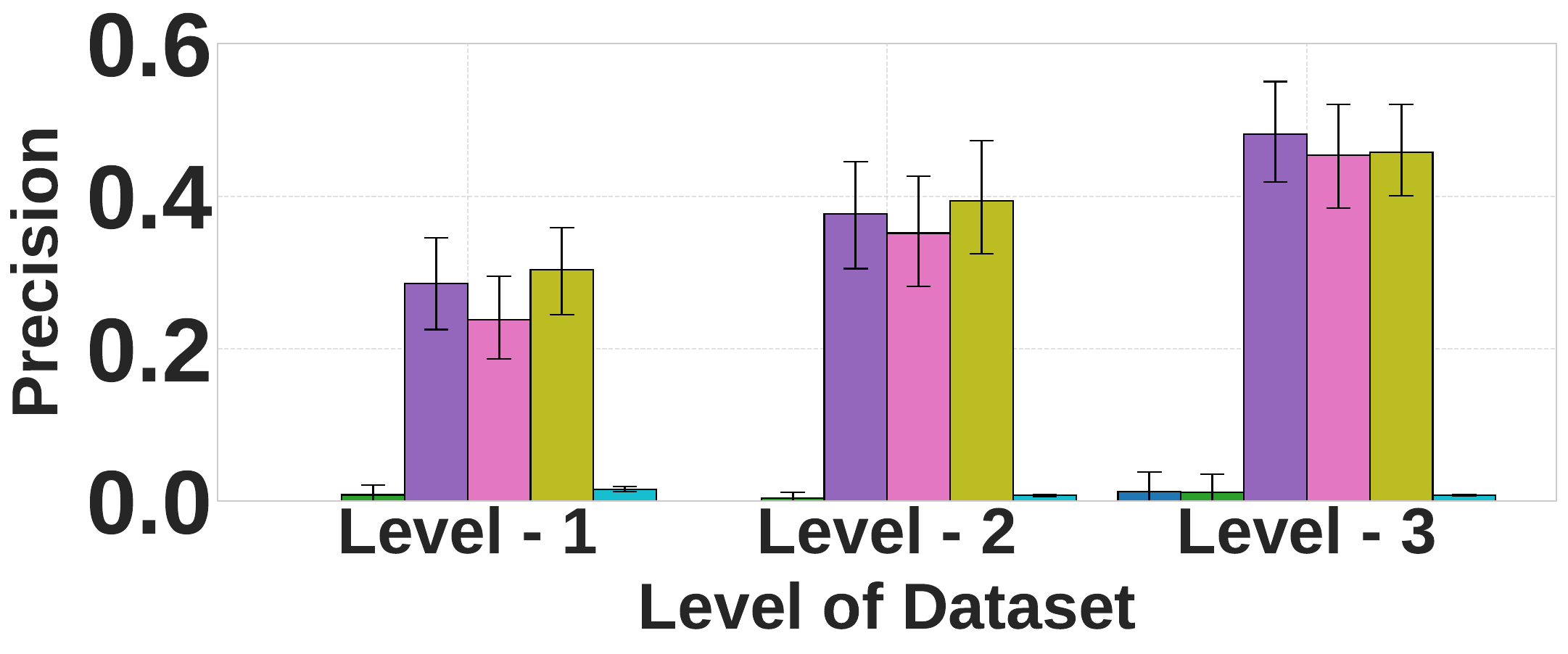}} 
    \hfill
    \subfloat[Recall - Function Selection]
    {\includegraphics[width=0.25\textwidth]{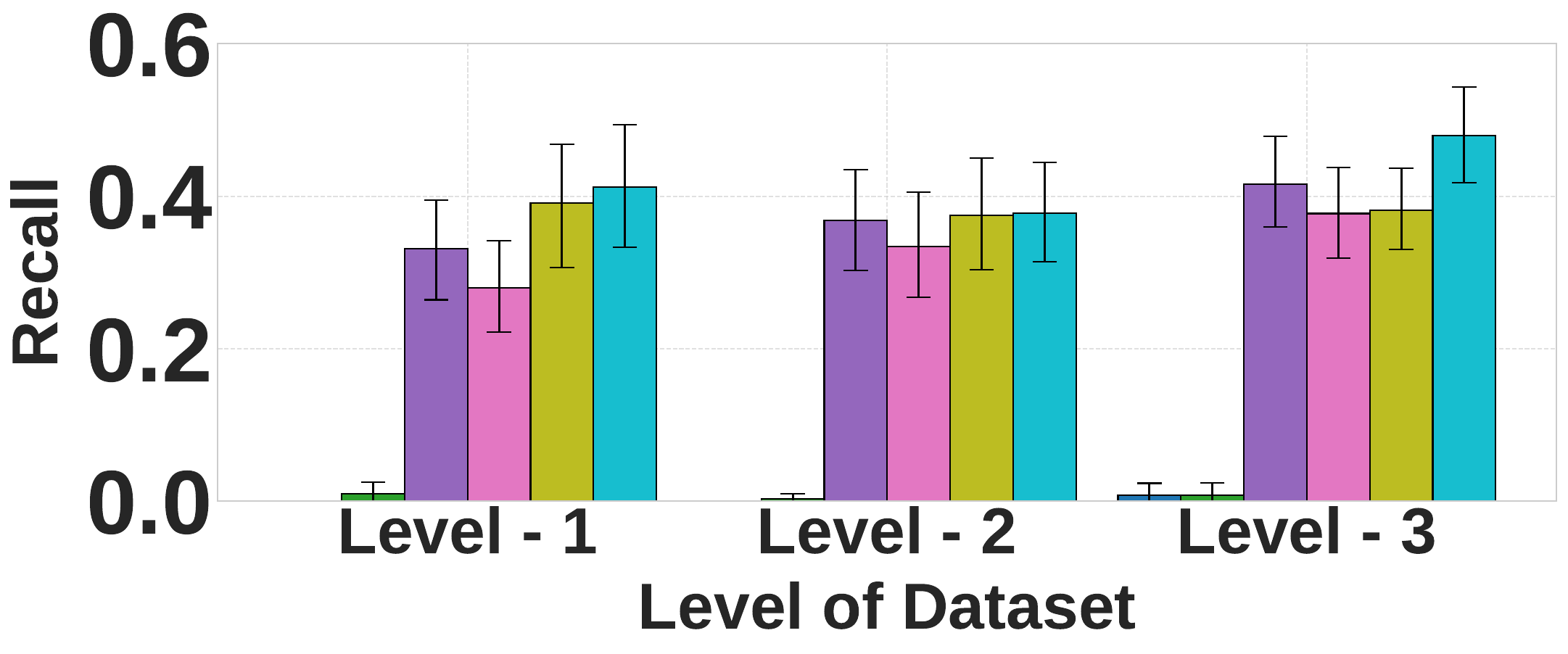}}
    \hfill
    \subfloat[Precision - Data Dependency]
    {\includegraphics[width=0.25\textwidth]{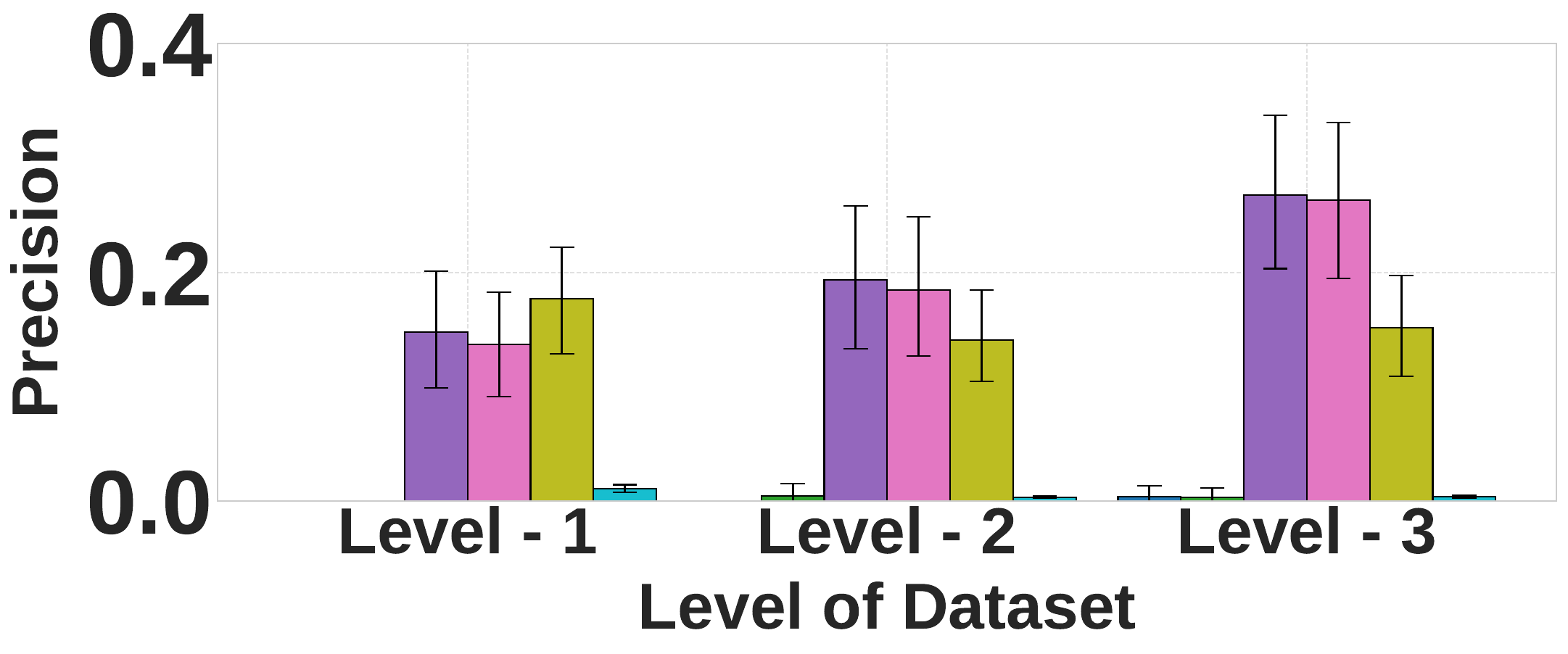}} 
    \hfill
    \subfloat[Recall - Data Dependency]
    {\includegraphics[width=0.25\textwidth]{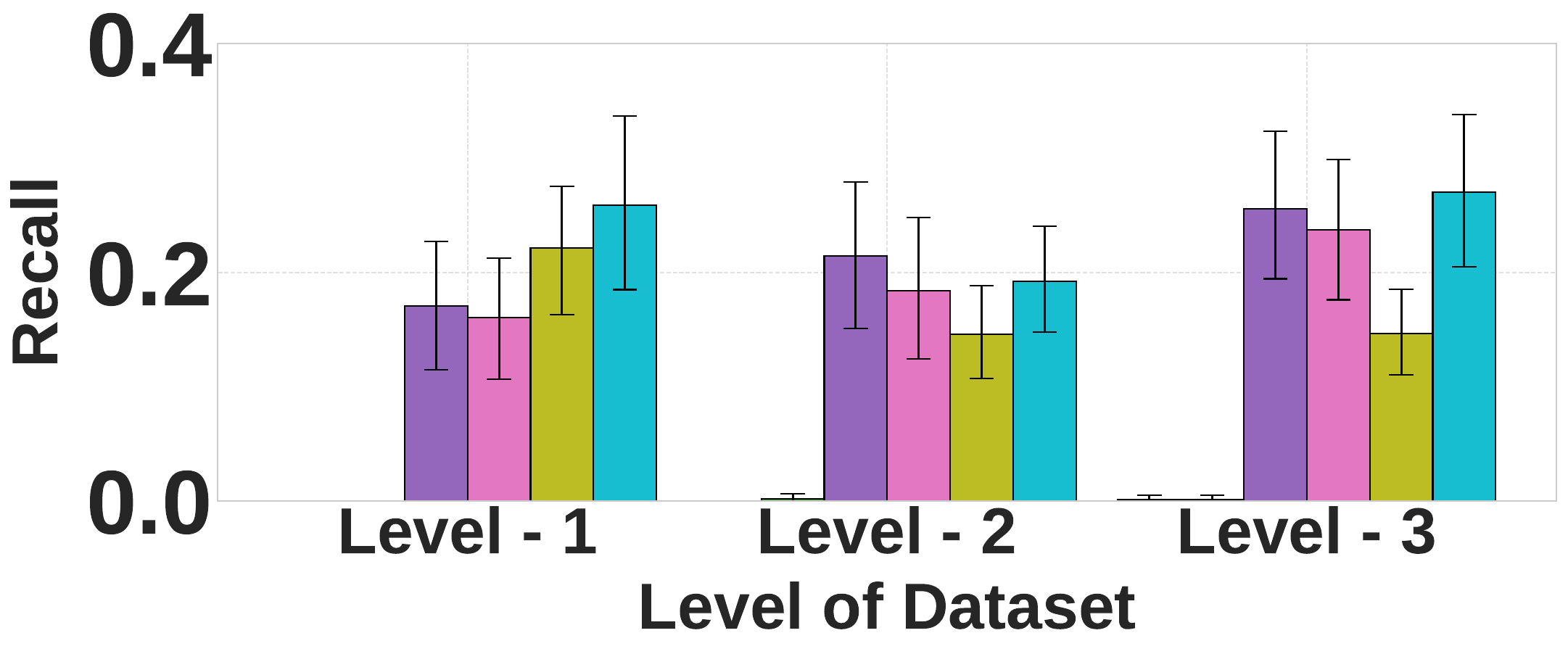}} 
    \vspace{-10pt}
    \hfill
    \subfloat{%
        \includegraphics[width=1\textwidth]{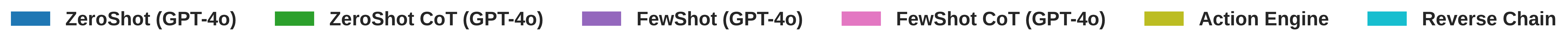}%
    }
    \vspace{-5pt}
    \caption{Precision and recall comparison for function selection and data dependency tasks. Reverse Chain shows high recall but low precision, reflecting a tendency to over-select, while Action Engine and FewShot (GPT-4o) offer more balanced performance.}

    \label{fig:top-down-ablation}
\end{figure*}

The Reverse Chain method, while conceptually promising as a top-down approach, demonstrates a mixed performance profile in our experiments. Despite its theoretical potential for higher accuracy, the results indicate that it is significantly less effective in terms of F1 scores. As shown in Figure \ref{fig:top-down-ablation}, this is primarily due to its low precision across all categories and levels, highlighting a tendency to select extra nodes and increase false positives. However, the method exhibits strong recall, particularly in Function Selection and Parameter Dependency tasks, where it often outperforms or matches alternative methods such as the Action Engine and FewShot (GPT-4o) approaches. This suggests that the Reverse Chain method effectively identifies relevant functions and dependencies, capturing the ground truth with high sensitivity.

\begin{figure*}[h]
    \vspace{-15pt}

    \centering
    \subfloat[F1 Score - Function Selection]{%
        \includegraphics[width=0.3\textwidth]{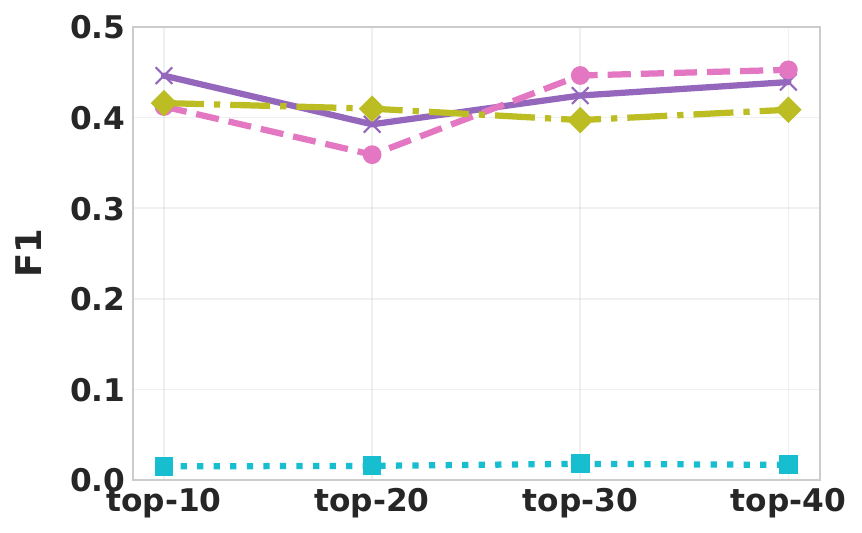}%
    }
    \hfill
    \subfloat[F1 Score - Data Dependency]{%
        \includegraphics[width=0.3\textwidth]{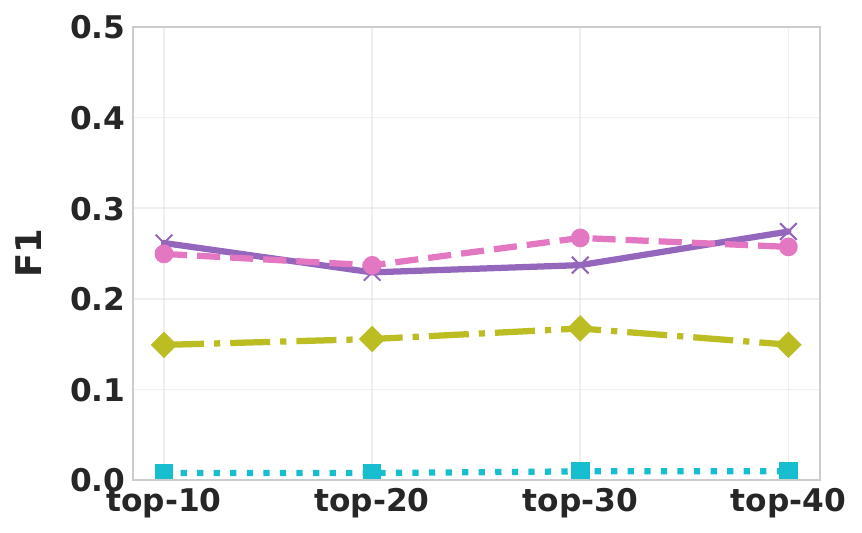}%
    }
    \hfill
    \subfloat[LCS score - Topological Order]{%
        \includegraphics[width=0.3\textwidth]{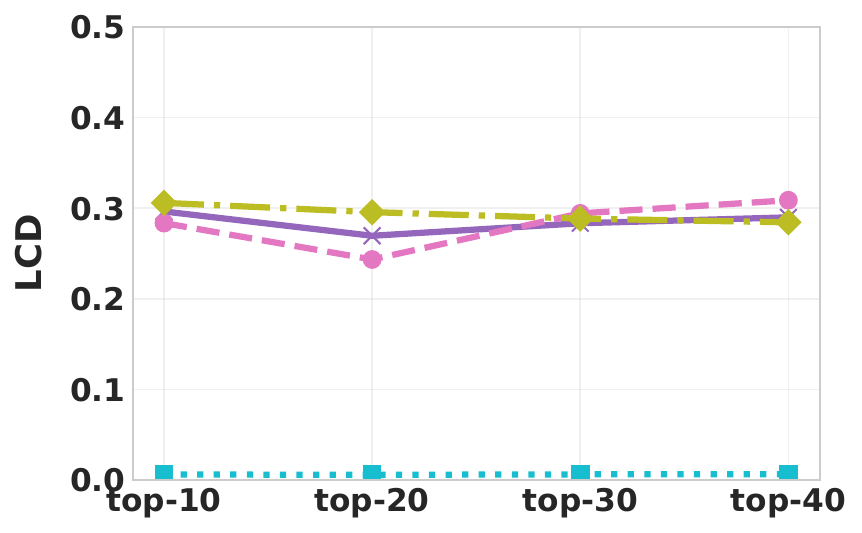}%
    }
    \vspace{-10pt}
    \hfill
    \subfloat{%
        \includegraphics[width=0.7\textwidth]{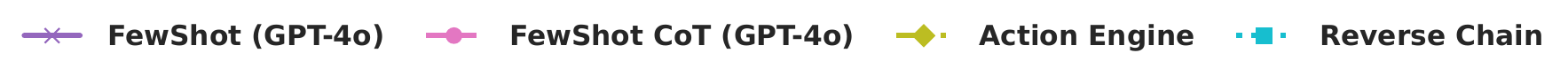}%
    }
    \vspace{-5pt}
    \caption{Impact of increasing Top-K candidate functions on workflow generation performance. }
    \label{fig:TopK-Impact_Perforance}
    \vspace{-15pt}
\end{figure*}
A key limitation stems from the evaluation strategy employed in the original Reverse Chain method, which relies on an LLM to semantically assess function correctness. While LLM-based evaluation is effective for nuanced semantic understanding, it does not adequately account for strict syntactic and logical correctness, which is crucial in FaaS workflows. Additionally, the preprocessing phase findings revealed syntactic and logical errors in some ground truth---compounding the method's precision issues. The disparity between low precision and high recall underscores a fundamental tradeoff. While the method excels at identifying relevant elements, its over-selection of irrelevant nodes reduces its overall effectiveness, as reflected in its low F1 scores. These findings suggest that the top-down approach, such as the Reverse Chain method, is potentially effective. However, it requires further refinement to improve its specificity, precision, and robustness in handling syntactic and logical correctness.

\input{Table_1}

\subsubsection{The Impact of Choosing Top-K Functions}
 We investigate the effect of increasing the number of Top-K functions provided to the LLM for function selection and evaluate its impact on overall performance with GPT-4o. As shown in Figure \ref{fig:TopK-Impact_Perforance}, we assess the results across three key metrics: Function Selection Accuracy, Data Dependency Accuracy, and Topological Order Accuracy. Our findings indicate that increasing the number of candidate functions does not lead to a significant improvement in accuracy scores across these metrics.

However, a more nuanced effect emerges when examining the Pass Rate shown in Table \ref{tab:pass_rate}, which measures the ability of the LLM to generate a syntactically valid and executable workflow. Our analysis reveals that as the number of Top-K candidate functions increases, the Pass Rate declines, indicating that an excessive number of function choices may introduce ambiguity and complexity that hinder the ability of LLMs to generate fully structured workflows. In contrast, \name, which systematically constructs workflows following the conceptual DAG creation process with the compiler, consistently achieves better accuracy in syntactic correctness regardless of the number of Top-K functions considered.

These findings highlight a fundamental limitation of direct LLM-based workflow generation. While expanding the candidate function set does not meaningfully enhance selection accuracy, it compromises syntactic and execution correctness. In contrast, systematic approaches like \name offer a more structured and reliable alternative, ensuring syntactically correct and executable workflows. These methods are better suited for applications requiring strict correctness and scalability in handling complex execution requirements.

\subsubsection{Parameter Accuracy}
To confirm the validity of our results, we conducted an ablation study to further investigate the robustness of dependency predictions. Specifically, we focused on Data Dependency Accuracy to assess how well different methods capture dependencies under controlled conditions. Our analysis revealed that the dataset contains multiple valid solution paths, which can introduce inconsistencies in evaluation and affect the robustness of reported accuracy scores. To mitigate this issue and isolate the true effectiveness of different methods, we performed an additional evaluation by providing the ground truth function set to the three highest-performing methods: FewShot (GPT-4o), FewShot CoT (GPT-4o), and \name.

\begin{figure}[t]
    
    \centering
    \subfloat{
        \includegraphics[width=0.7\columnwidth]{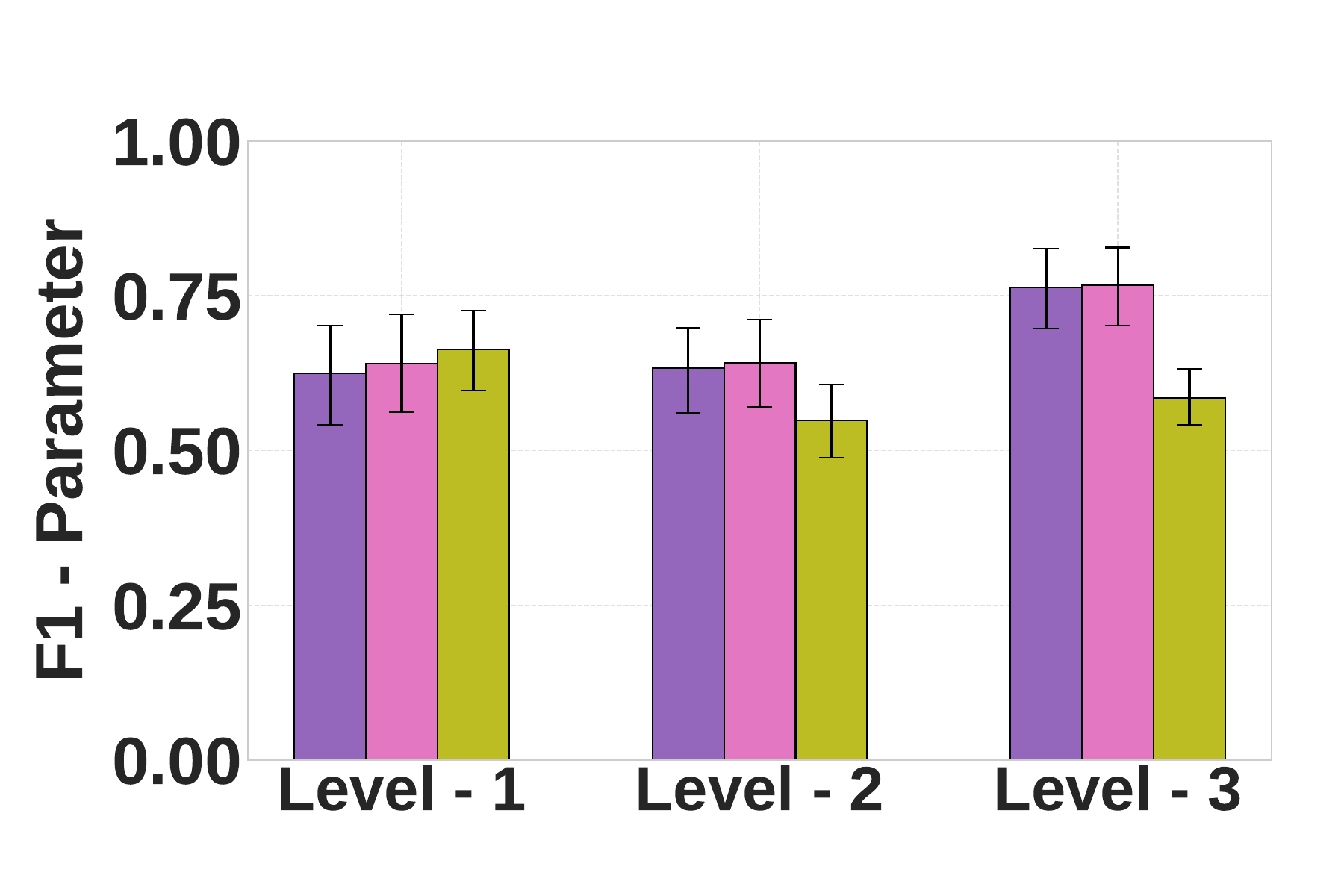}
    }
    \vspace{-15pt}
    \hfill
    \subfloat{%
        \includegraphics[width=1\columnwidth]{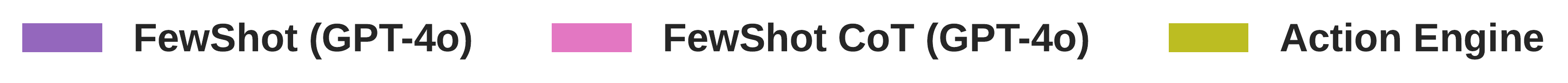}%
    }
    \caption{Data dependency F1 scores across the top three methods (FewShot, FewShot CoT, and \name) after constraining function selection to the ground truth function set. }
    \vspace{-15pt}
    \label{ablation:param} 
\end{figure}

By constraining function selection to the predefined ground truth and evaluating only the ability to identify correct data dependencies (Figure \ref{ablation:param}), we observed a nearly threefold increase in accuracy across all methods. This significant improvement underscores the critical role of accurate function selection in workflow evaluation. However, despite the absolute increase in accuracy, the overall ranking of methods remained unchanged, reinforcing the credibility of our primary findings.

%% file: Table_1.tex
\begin{table}[h]
    \centering
    \captionsetup{justification=centering,singlelinecheck=false}  
    \begin{tabular}{l|cccc}
        \hline
        Method & top-10 & top-20 & top-30 & top-40 \\ 
        \hline
        ActionEngine & 100 & 100 & 100 & 100 \\
        FewShot & 91 & 99 & 84 & 81 \\
        FewShot-CoT & 99 & 99 & 98 & 92 \\
        ReverseChain & 59 & 59 & 57 & 55 \\
        \hline
    \end{tabular}
    \caption{The pass rate (\%) comparison across different Top-K function selections, which measures the ability of the LLM to generate syntactically valid and executable workflows as the number of Top-K candidate functions increases.}
    \label{tab:pass_rate}
\end{table}

%% file: limitation.tex
\section{Limitation and Future work}
\label{sec:limitation_futurework}

Despite the \name's ability to generate relatively accurate FaaS workflows while achieving language/platform neutralization, there remains a non-negligible likelihood of inaccurate topological features, emphasizing the importance of further improving accuracy and reliability. As such, we consider the current version of \name as an ``assistant'' to cloud developers that, in practice, can significantly reduce the lead time required to generate FaaS workflows. However, developer intervention and verification remain essential to ensure workflow correctness and optimal execution. Common situations where such oversight is critical include incorrect workflow structures, misaligned data dependencies, and incorrect function selections, each of which can compromise workflow integrity and efficiency. 

Since this work focuses exclusively on workflow generation performance, we do not evaluate the performance of workflow execution, as it is highly dependent on the specific workflow orchestrator and FaaS runtime environment used. Incorporating such execution-level performance analysis is considered out of the scope of this study.

To mitigate these limitations and move toward more autonomous and reliable workflow generation, we propose several future enhancements. First, we hypothesize that enhancing data dependency, currently the primary bottleneck, can significantly improve the accuracy of workflow generation. Additionally, integrating LLM-generated test cases into the post-generation validation process offers a promising avenue. These test cases can automatically detect and correct structural flaws and inconsistencies in data flow, supporting both correctness verification and performance profiling.

Building on this foundation, we envision an automated feedback loop where LLMs not only validate workflows but also generate targeted test scenarios and automatically correct or improve the workflow. This integration would address broader software development needs, including performance, efficiency, and scalability. Furthermore, we plan to incorporate formal reliability guarantees into the generation process \cite{yang2025convex}, enhancing trust in the robustness and resilience of workflows across diverse and potentially uncertain FaaS environments.

To further support real-world applicability, integrating Action Engine into a FaaS workflow orchestrator on industry-standard cloud platforms (e.g., AWS Step Functions~\cite{AWS-step-functions} or Google Cloud Composer~\cite{GoogleCloudComposer}), a new compiler needs to be implemented. Moreover, FaaS continues to face inherent challenges related to managing intermediate data. Specifically, functions might need to store intermediate data in cloud storage (e.g., AWS S3~\cite{aws_s3}). Rather than passing the actual data, which may be infeasible due to data size limitations, functions return a reference to the stored data (e.g., a URL to S3). As a result, downstream functions must be aware of this practice and capable of retrieving data using these references. This issue may add more complexity for the LLM to recognize and correctly map the data dependency.

Lastly, the Action Engine’s reliance on GPT4o for key decisions like function selection and data dependency management introduces variability when considering alternative LLMs. In particular, open-source models may require significant prompt engineering or tuning to match the performance of proprietary models, due to differences in model size and training approaches (e.g., raw, instruct, or chat-optimized). 

%% file: conclsn.tex
\section{Conclusion}
\label{sec:conclusion}
This research presents \name, a novel system that leverages tool-augmented LLMs for automatic FaaS workflow generation. By integrating function selection, data dependency management, and structured workflow execution, \name significantly reduces the need for manual intervention and specialized expertise, making FaaS-based application development more accessible and efficient.

Our evaluations demonstrate that \name achieves comparable performance to few-shot learning methods, while maintaining a language and platform-independent design. Through rigorous benchmarking, we highlight key factors that influence workflow generation accuracy, including dataset ambiguity, function repository quality, and evaluation methodology limitations. Notably, our findings underscore the need for evaluation frameworks that account for multiple valid solutions, rather than relying solely on a rigid and binary topological correctness workflow.
